\documentclass[sigconf,natbib=true,anonymous=false]{acmart}

\AtBeginDocument{%
  \providecommand\BibTeX{{%
    \normalfont B\kern-0.5em{\scshape i\kern-0.25em b}\kern-0.8em\TeX}}}

\usepackage{booktabs}
\usepackage{xspace}
\usepackage{tcolorbox}
\usepackage{xcolor} 
\usepackage{tikz}
\usepackage{caption}
\usepackage{fontawesome}
\usepackage{customdice}
\usepackage{diagbox}

\usepackage[show]{chato-notes}

\tcbuselibrary{skins, breakable, xparse}

\definecolor{gold}{RGB}{212, 175, 55}
\definecolor{correct}{RGB}{150, 200, 50}

\copyrightyear{2024}
\acmYear{2024}
\setcopyright{rightsretained}
\acmConference[SIGIR '24]{Proceedings of the 47th International ACM SIGIR Conference on Research and Development in Information Retrieval}{July 14--18, 2024}{Washington, DC, USA}
\acmBooktitle{Proceedings of the 47th International ACM SIGIR Conference on Research and Development in Information Retrieval (SIGIR '24), July 14--18, 2024, Washington, DC, USA}
\acmDOI{10.1145/3626772.3657834}
\acmISBN{979-8-4007-0431-4/24/07}

\begin{document}

\title{The Power of Noise: Redefining Retrieval for RAG Systems}



\author{Florin Cuconasu$^*$}
\orcid{0009-0008-9768-1047}
\email{cuconasu@diag.uniroma1.it}
\affiliation{%
  \institution{Sapienza University of Rome}
  \city{Rome}
  \country{Italy}
}


\author{Giovanni Trappolini$^*$}
\orcid{0000-0002-5515-634X}
\email{trappolini@diag.uniroma1.it}
\affiliation{%
  \institution{Sapienza University of Rome}
  \city{Rome}
  \country{Italy}
}

\author{Federico Siciliano}
\orcid{0000-0003-1339-6983}
\email{siciliano@diag.uniroma1.it}
\affiliation{%
  \institution{Sapienza University of Rome}
  \city{Rome}
  \country{Italy}
}

\author{Simone Filice}
\orcid{0009-0002-6735-9950}
\email{filice.simone@gmail.com}
\affiliation{%
  \institution{Technology Innovation Institute}
  \city{Haifa}
  \country{Israel}
}

\author{Cesare Campagnano}
\orcid{0000-0002-8362-274X}
\email{campagnano@di.uniroma1.it}
\affiliation{%
  \institution{Sapienza University of Rome}
  \city{Rome}
  \country{Italy}
}

\author{Yoelle Maarek}
\orcid{0000-0003-3160-7115}
\email{yoelle@yahoo.com}
\affiliation{
  \institution{Technology Innovation Institute}
  \city{Haifa}
  \country{Israel}
}

\author{Nicola Tonellotto}
\orcid{0000-0002-7427-1001}
\email{nicola.tonellotto@unipi.it}
\affiliation{%
  \institution{University of Pisa}
  \city{Pisa}
  \country{Italy}
}

\author{Fabrizio Silvestri}
\orcid{0000-0001-7669-9055}
\email{fsilvestri@diag.uniroma1.it}
\affiliation{%
  \institution{Sapienza University of Rome}
  \city{Rome}
  \country{Italy}
}



\renewcommand{\shortauthors}{Cuconasu and Trappolini, et al.}

\begin{abstract}
Retrieval-Augmented Generation (RAG) has recently emerged as a method to extend beyond the pre-trained knowledge of Large Language Models by augmenting the original prompt with relevant passages or documents retrieved by an Information Retrieval (IR) system. 
RAG has become increasingly important for Generative AI solutions, especially in enterprise settings or in any domain in which knowledge is constantly refreshed and cannot be memorized in the LLM. We argue here that the retrieval component of RAG systems, be it dense or sparse, deserves increased attention from the research community, and accordingly, we conduct the first comprehensive and systematic examination of the retrieval strategy of RAG systems. 
We focus, in particular, on the type of passages IR systems within a RAG solution should retrieve. 
Our analysis considers multiple factors, such as the relevance of the passages included in the prompt context, their position, and their number.
One counter-intuitive finding of this work is that the retriever's highest-scoring documents that are not directly relevant to the query (e.g., do not contain the answer) negatively impact the effectiveness of the LLM.
Even more surprising, we discovered that adding random documents in the prompt improves the LLM accuracy by up to 35\%.
These results highlight the need to investigate the appropriate strategies when integrating retrieval with LLMs, thereby laying the groundwork for future research in this area.\footnote{The code and data are available at \href{https://github.com/florin-git/The-Power-of-Noise}{github.com/florin-git/The-Power-of-Noise}}

\def\thefootnote{*}\footnotetext{These authors contributed equally to this work.}
\end{abstract}







\maketitle



%
%
%
%
%
%
%
%
%
%
%
%
%
%
%
%
%
%
%
%
%
%
%
%
%
%
%
%
%
%
%
%
%
%
%
%

\section{Introduction}

Large Language Models (LLMs) \cite{brown2020language} have demonstrated unprecedented proficiency in various tasks, ranging from text generation and complex question answering \cite{open-llm-leaderboard}, to information retrieval~(IR) tasks \cite{kenton2019bert,yates-etal-2021-pretrained}.
However, LLMs have limitations in the handling of long contexts \cite{vaswani2017attention}, a constraint that leads to an increased reliance on their pre-trained knowledge. 
This limitation not only confines their ability to effectively manage extended discourse, such as in books or long conversations, but also increases the probability of generating hallucinations, instances for which the model produces factually incorrect or nonsensical information \cite{roller-etal-2021-recipes}. 
To improve the accuracy of responses generated by LLMs, Retrieval-Augmented Generation (RAG) has emerged as a promising solution \cite{lewis2020retrieval}. 
RAG is primarily designed to improve factual accuracy by providing the model access to auxiliary information, thereby augmenting the original prompt with information not necessarily memorized in the LLM. A key benefit of this approach is that it helps ground the prompt with relevant information that might help the LLM generate more accurate answers at inference time.
At their core, RAG systems consist of two fundamental components: a \textit{retriever} and a \textit{generator}. The retriever is responsible for invoking an external IR system (dense and/or sparse) and feeding the selected results to a generator component. 

This study focuses on the IR aspect of RAG, posing the following research question:
\emph{``What characteristics are desirable in a retriever to optimize prompt construction for RAG systems? Are current retrievers ideal?"}.
We focus on the three main types of documents (or passages\footnote{We interchangeably use here the terms ``passage" or ``document" to represent the indexing/retrieval unit of the IR system.}) that a retriever can return: \textit{relevant}, \textit{distracting}, and \textit{random}. 
\textit{Relevant} documents contain pertinent information that either directly answers or might inform the query.
\textit{Distracting} documents, while not directly answering the query, are semantically or contextually linked to the topic. For instance, if one asks for the color of Napol\'{e}on's horse, a passage describing the color of Jos\'{e}phine de Beauharnais' (Napol\'{e}on's first wife) horse, while not containing the right information, would be highly related.
\textit{Random} documents
have no relation whatsoever to the query and can be seen as a kind of informational noise within the retrieval process. 
One of the key goals of our study is to determine the role of each type of document and the relative value they bring to the LLM effectiveness. In particular, we verify whether there is a need to revisit some of the commonly accepted assumptions in IR systems when used in the context of LLMs.
The main contributions of our work are the following:
\begin{enumerate}
\item We conduct the first comprehensive study examining the impact of the type of retrieved documents in RAG on the LLM effectiveness.
\item We propose retrieval RAG heuristics that leverage the unexpected results of this study.
\item We release all associated code and data to the community to encourage further research.
\end{enumerate}

\section{Related Works}

\subsection{Generative Language Models}

\looseness -1 The inception of the modern LLM era can be traced back to the seminal paper titled ``Attention Is All You Need" \cite{vaswani2017attention}. 
This work introduced the transformer architecture, a framework that adopts an attention mechanism instead of recurrent layers, enabling the model to capture global dependencies within the data. 
%
The following year, BERT (Bidirectional Encoder Representations from Transformers) \cite{kenton2019bert} offered a significant improvement over the state-of-the-art via a novel bidirectional, unsupervised language representation. 
%
The evolution of transformer-based models continued with the development of the Generative Pre-trained Transformer (GPT) \cite{radford2018improving}. 
Its successor, GPT-2 \cite{radford2019language}, expanded upon this foundation with a larger scale model and demonstrated improved performance across a variety of language tasks without task-specific training. 
The subsequent iteration, GPT-3 \cite{brown2020language}, represented a further enhancement in model scale and capabilities, particularly in the realm of few-shot learning. 
Finally, recent times have seen a surge in the production of large, publicly available language models.
Several actors have released their models, most notably, Llama \cite{touvron2023llama,touvron2023llama2}, Falcon \cite{almazrouei2023falcon}, Mosaic MPT \cite{mosaicml2023introducing}, and Phi \cite{li2023textbooks,javaheripi2023phi}.
There are also versions of these models that have been fine-tuned on specific languages \cite{Garrachonr_2023,jphme_2023,Chinese-LLaMA-Alpaca,santilli2023camoscio,fauno}.
The proliferation and quality of these models are expanding the range of tasks and the vision they address \cite{wang2024oop,xie2024osworld,prompt2os}.

\subsection{Information Retrieval}

Foundational information retrieval methodologies, such as the Vector Space Model 
and the TF-IDF scoring \cite{salton1983introduction}
introduced in the 1980s are the basis for quantifying textual similarity. 
These retrieval methods are characterized by their use of high-dimensional and sparse feature vectors and have been essential in developing a full generation of IR systems. BM25 represents the most famous current iteration \cite{robertson2009probabilistic}. 
A significant evolution in IR is the introduction of dense retrievers, which emerged from advancements in deep learning; they utilize low-dimensional dense vectors for textual representation, and allow to capture semantic relationships. This is in contrast to traditional IR methods (referred to as sparse in opposition to dense), which typically rely on lexical match and struggle with semantic match~\cite{manning2008term}. In the last few years, dense methods such as DPR \cite{karpukhin2020dense} and others \cite{khattab2020colbert, izacard2021unsupervised} have demonstrated that they can compete with sparse methods.


\subsection{Retrieve and Generate}

RAG introduces a new approach in AI, combining the strengths of both retrieval-based and generative models. 
The concept of RAG was coined and popularized in \cite{lewis2020retrieval}, which introduced a model that combines a dense passage retriever with a sequence-to-sequence model, demonstrating substantial improvements in knowledge-intensive tasks. Similar methods/variations have also been proposed concurrently or soon after, such as \cite{guu2020retrieval,borgeaud2022improving,asai2023selfrag,ke2024bridging,rraml}; see \cite{mialon2023augmented} for a survey on augmented language models.
Researchers and practitioners have recently started to explore these RAG systems' inner workings.
Notably, \cite{sauchuk2022role,mmndb} analyzed the impact of different types of documents on cascading IR/NLP systems.
Other works have tried to study how attentive transformers are to their input \cite{sun2021long,khandelwal2018sharp,liu2023lost,ram2023context,lu-etal-2022-fantastically}.
\cite{behnamghader2022can} studied the effect of the retriever's similarity metric, which was found to be insufficient for reasoning.
In \cite{xie2023adaptive,koopman-zuccon-2023-dr}, authors analyzed LLM's receptiveness to external evidence against internal memory.
In \cite{zuccon_hallucinates}, they test the model's (in)ability to ground references.

In this paper, we want to provide the first comprehensive analysis of the implications of using a retriever module in a RAG system, studying the impact of several key factors, like the type, number, and position of documents that should augment the prompt to the LLM.

\section{RAG}
\label{sec:over_RAG}

In this paper, we explore the application of RAG in the context of Question Answering, arguably its most popular application.

\subsection{Open-Domain Question Answering}

Open-domain Question Answering (OpenQA) refers to the task of developing systems capable of providing accurate and contextually relevant answers to a broad range of questions posed in natural language without limitations to specific domains or predefined datasets. 
In general, we want to find an answer $\mathcal{A}$ to a query $q$.
To do so, we draw information from a corpus of documents $\mathcal{D} = \{d_1, d_2, \dots, d_n\}$, which is usually assumed to be large in size.
A prevalent approach for this task involves a two-step architecture, typically comprising a retriever and a reasoner (typically a generator). This methodology addresses the inherent complexities of OpenQA by dividing the process into distinct phases: 
first finding the appropriate set of documents that can potentially address the query and then synthesizing an answer, which can be consumed by the user of the QA system.

\subsection{Retriever}\label{sec:over_retr}
The retriever plays a critical role in the OpenQA task. 
Its goal is to find a sufficiently small subset of documents $\mathcal{D}_r$ to allow the reasoner to answer the query correctly.
Among the various retrieval methodologies, the use of a dense retriever has gained prominence due to its effectiveness in handling semantic matches. Dense retrieval requires transforming textual data into vector representations, which is typically achieved with a neural network, often a transformer-based encoder, like BERT \cite{kenton2019bert}. 
The dense retriever processes both the query $q$ and potential source documents to generate corresponding embeddings $\vec{q}$ for the query and $\vec{d_i}$ for each document $d_i \in \mathcal{D}$.
The embedding process can be represented as:
$$
\vec{q} = Encoder_q(q); \; \vec{d_i} = Encoder_d(d_i)
$$
where $Encoder_q$ and $Encoder_d$ are neural network-based encoders, potentially sharing weights or architecture, designed to map the textual data into a vector space.
Once the embeddings are generated, the retrieval process involves computing the similarity between the query embedding and each document embedding. The most common approach is to use dot product \cite{DPR}, defined as: $s(q, d_i) = \vec{q} \cdot \vec{d_i}$.
This score quantifies the relevance of each document to the query by measuring their similarity in the embedded vector space, with higher scores indicating greater relevance. According to these scores, the top-ranked documents are selected for further processing in the generator component.

\subsection{Reasoner} 

The second step involves a generator component in charge of synthesizing an answer, typically implemented via an LLM.
Generative language models operate by predicting the probability distribution of the next token, given the previous tokens. 
For a given sequence of words $w_1, w_2, \dots, w_n$, a generative language model aims to maximize the likelihood of this sequence, expressed using the chain rule of probability:

\begin{equation*}
P (w_1, w_2, \dots, w_n) = \prod_{i=1}^{N} P(w_i | w_1, w_2, \dots, w_{i-1})
\end{equation*}
where $P(w_i | w_1, w_2, \dots, w_{i-1})$ is the conditional probability of the word $w_i$ given the preceding sequence of words $w_1, w_2, \dots, w_{i-1}$.
In RAG, the generative language model takes a query $q$ and the retrieved documents $\mathcal{D}_r$ as input and generates a response by sequentially predicting the next token in the sequence.
More formally, 
\begin{equation*}
    P_{rag} (y|q) \approx \prod_{i}^{N} \sum_{d \in \mathcal{D}_r} p_{\eta} (d | q)  p_{\theta} (y_i | q, d, y_{1:i-1}),
\end{equation*}
where $p_{\eta} (d | q)$ is the retrieval component that provides a (truncated) probability distribution for the top-scoring documents, and $p_{\theta} (y_i | q, d, y_{1:i-1})$ is a probability distribution parameterized by $\theta$ that generates a current token based on the previously generated tokens, the query, and the retrieved document; this role is filled by the LLM.
In the case of dense retrieval, the probability distribution for the top-scoring documents may assume a functional form of the kind $p_{\eta} (d | q) \propto \exp(\vec{q} \cdot \vec{d})$.
Given our formalization of the RAG task, we notice how the generative component $p_\theta$ depends on a given text, that is the query, and a dynamic text, that is the set of retrieved documents.
We study in the next two sections the impact of changing the set of retrieved documents on the generator and, consequently, the whole end-to-end system. In particular, we aim to find the best set of documents $\mathcal{D}_r$ that a retriever should feed the generator to maximize the system's effectiveness.

%

%
\section{Experimental Methodology}
\label{sec:exp_set}

In this section, we detail the experimental framework.
We start by describing the data used in the experiments and then discuss the type of documents that a retriever can return and pass to the LLM.


\subsection{Natural Question Dataset}

The Natural Questions (NQ) dataset \cite{nq} is a large-scale collection of real-world queries derived from Google search data. Each entry in the dataset consists of a user query and the corresponding Wikipedia page containing the answer. 
The NQ-open dataset \cite{nq-open}, a subset of the NQ dataset, differs by removing the restriction of linking answers to specific Wikipedia passages, thereby mimicking a more general information retrieval scenario similar to web searches. This open-domain nature significantly impacts our experimental design, particularly in the selection and categorization of documents.
Following the methodology of \citet{nq-open}, our primary source for answering queries is the English Wikipedia dump as of 20 December 2018. 
Consistently with the Dense Passage Retrieval (DPR) approach \cite{DPR}, each Wikipedia article in this dump was segmented into non-overlapping passages of $100$ words.
A significant challenge in open-domain question answering is the potential temporal mismatch between the Wikipedia dump and the question-answer pairs in the dataset, which can lead to missing answers in the dataset, as highlighted in the AmbigQA study \cite{ambigqa}.
To mitigate this, 
we integrated the gold documents from the original NQ dataset into our Wikipedia document set. 
Given the open-domain nature of our task, there may be additional documents \emph{relevant} to the query, i.e., containing the answer, but we will \emph{not} consider them as \emph{gold}.
The final dataset comprises $21,035,236$ documents, with $72,209$ queries in the train set and $2,889$ in the test set.


\subsection{Types of Documents} \label{doc_types-section}
In our study, we categorize documents into four distinct types, each represented by a unique symbol, based on their relevance and relationship to the queries:

\paragraph{\faStar \ Gold Document} 
The gold document, identified by \faStar, refers to the original context in the NQ dataset, specifically the passage of a Wikipedia page containing the answer and contextually relevant to a given query.

\paragraph{\faChain \ Relevant Documents} 
Denoted by \faChain, relevant documents are passages that, akin to the gold document, contain the correct answer and are contextually useful for answering the query. They provide additional sources of information that are correct and pertinent to the query. 
Notably, the gold document is a relevant document.

\paragraph{\faChainBroken \ Distracting Documents}
Symbolized by \faChainBroken, distracting documents are semantically similar to the query but do not contain the correct answer. They serve a crucial role in evaluating the generator's proficiency in discerning between relevant and non-relevant information. In practice, these are the top-scoring retrieved documents that are not relevant.

\paragraph{\dice{6} \ Random Documents}
Indicated by \dice{6}, random documents are neither related to the query nor contain the answer. They are instrumental in assessing the model's ability to handle completely unrelated information. In practice, in our tests, we will randomly sample these documents from the corpus.
\\ \\
In our analysis, the entire set of documents fetched by the retriever is represented by the symbol \faFileTextO. This possibly encompasses all document types — gold, relevant, distracting, or random — and serves to discuss the retrieval output in a generalized manner without specifying individual document categories.


\subsection{Document Retrieval}
Our methodology utilizes a two-step approach in line with a typical RAG setting, as explained in Section \ref{sec:over_retr}. 
As the first component, our experiments use \textit{Contriever} \cite{izacard2021unsupervised}, a BERT-based dense retriever, as the default retriever. It is trained without supervision using a contrastive loss. 
To enhance the efficiency of similarity searches within our corpus, comprising about 21 million documents, we also employ the FAISS IndexFlatIP indexing system \cite{douze2024faiss}. 
The embedding of each document and query is obtained by averaging the hidden state of the last layer of the model.

\subsection{LLM Input}
Upon receiving a query, the retriever selects the top-$k$ documents from the corpus according to a given similarity measure. 
These documents, in conjunction with the task instruction and the query, constitute the input for the LLM to generate a response.
The NQ-open dataset was structured to include only those queries whose answers consist of no more than five tokens \cite{nq-open}. 
Consequently, the LLM is tasked with extracting a query response, confined to a maximum of five tokens, from the provided documents. The input is encoded into a prompt, whose template is shown in Figure \ref{fig:prompt_only_gold}, beginning with the task instruction, presented in italics for clarity. This is followed by the \emph{context}, which comprises the selected documents followed by the query string. This prompt design aligns with the methodological approach outlined in \cite{liu2023lost}.
\begin{figure}[htbp]
\begin{tcolorbox}[title=LLM Input - Only Gold \faStar]
\emph{You are given a question and you MUST respond by EXTRACTING the answer (max 5 tokens) from one of the provided documents. If none of the documents contain the answer, respond with NO-RES.}

\textbf{Documents:}

\textcolor{gold}{\textbf{Document [3]}}(Title: Millennium Falcon) Han Solo won the Millennium Falcon from \textcolor{gold}{Lando Calrissian} in the card game sabacc...
\\ \\
\textbf{Question:} who owned the millennium falcon before han solo

\textbf{Answer:} \textcolor{red}{Han Solo}
\end{tcolorbox} 
\caption{Example LLM input with an erroneous output, highlighted in \textcolor{red}{red}. The input consists of an \emph{italicized task instruction}, followed by the context (documents), and the query. The LLM's response is marked under `Answer'. The \textcolor{gold}{gold} color highlights both the gold document and the correct answer, ``Lando Calrissian'', indicating the expected source and content of the accurate response.}

\label{fig:prompt_only_gold}
\end{figure}
While the composition of the context will vary according to the single experiment, the instruction will always be placed at the beginning of the prompt and the query always at the end.

\subsection{LLMs Tested}

We consider several LLMs in our experiments.
Consistently across all models, we adopt a greedy generation approach with a maximum response length of 15 tokens.
Acknowledging the constraints imposed by memory and computational resources, we have implemented a model quantization strategy, reducing all models to a 4-bit representation.
Besides the above prompt, the models are not provided with additional exemplars for few-shot learning, which, while of interest, is outside the scope of this paper.
We conduct tests on both the \emph{base} and the \emph{instruct} versions of the LLMs. However, we only report on the latter, as while the behavior is consistent across both, the instruct versions demonstrate superior performance.
\begin{itemize}
\item \emph{Llama2.} 
The 7B parameters version of the Llama2 family \cite{touvron2023llama2} shows state-of-the-art performance on most downstream tasks compared to models of the same size. It was trained with a 4096 tokens context window and uses multi-query attention \cite{shazeer2019mqa}.
\item \emph{Falcon.} 
Falcon 7B, the smallest model of the Falcon series, \cite{almazrouei2023falcon} was trained on the RefinedWeb dataset \cite{penedo2023refinedweb}, a large, filtered, and deduplicated corpus. Similarly to Llama2, it uses multi-query attention, with a context length of 2048 tokens.
\item \emph{Phi-2.} 
This is the smallest model used in this work (2.7B parameters). Despite its modest size, it achieves performance comparable to the other models \cite{li2023textbooks,javaheripi2023phi}, thanks to its pre-training on ``textbook-quality'' data. It has a context window of 2048 tokens.
\item \emph{MPT.} 
This 7B parameters model uses ALiBi attention \cite{press2022alibi, mosaicml2023introducing} for a virtually unlimited context length. In our experiments, to leverage the model's full potential, we set the limit to 2048 tokens, i.e., the same used for the model's pre-training.
\end{itemize}
\subsection{Accuracy} \label{sec:eval}
The NQ-open dataset allows a range of potential answers for each query. 
Frequently, these answers are different variants of the same concept (e.g., ``President D. Roosevelt'' or ``President Roosevelt''), while in some cases, a single query may accept multiple distinct correct answers. To evaluate the accuracy of responses generated by LLMs, we use an assessment technique in line with \cite{same-accuracy,liu2023lost}.
This methodology examines whether at least one of the predefined correct answers is contained within the response produced by the LLM. 
%
We measure the correctness of the LLM’s responses as either accurate or inaccurate based on the presence of the answer in a binary fashion.
%
Nevertheless, this evaluation strategy is not without challenges. 
A principal issue arises in determining response correctness, particularly in instances involving date representations or varying phrasings conveying identical meanings. For example, if the LLM generates ``Roosevelt'' in response to a query where the established correct answer is ``President Roosevelt'', the response would be deemed incorrect under our current evaluation schema. Recognizing this limitation, we acknowledge the necessity for a more advanced analysis of answer variations, which we leave to future research.



\begin{table*}[!ht]
\caption{Accuracy results of the LLMs when evaluated with prompts composed of the gold document \faStar \ and a varying number of distracting \faChainBroken \ documents. The table illustrates how the inclusion of an increasing number of distracting documents affects LLM's performance. Scenarios where the prompt exceeded the model's input limit, leading to potential data truncation, are not included ( - ). All values \textit{not} marked with an asterisk * denote statistically significant changes from the gold-only document scenario [I, \faStar, Q] (first row), as determined by a Wilcoxon test (p-value < 0.01). Additionally, the closed-book accuracy scores for the models are as follows: Llama2 (0.1123), MPT (0.1205), Phi-2 (0.0488), Falcon (0.1083).}
\label{tab:rel_table}
\begin{tabular}{r|cccc|cccc|cccc@{}}
\toprule
           & \multicolumn{4}{c|}{\textbf{Far - [I, \faStar, \faChainBroken, Q]}}                                      & \multicolumn{4}{c|}{\textbf{Mid -  [I, \faChainBroken, \faStar, \faChainBroken, Q]}}                                     & \multicolumn{4}{c}{\textbf{Near -  [I, \faChainBroken, \faStar, Q]}}                                     \\ \midrule
\textbf{\# \faChainBroken} & \multicolumn{1}{c}{Llama2}          & \multicolumn{1}{c}{MPT}              & \multicolumn{1}{c}{Phi-2}            & \multicolumn{1}{c|}{Falcon}          & \multicolumn{1}{c}{Llama2}          & \multicolumn{1}{c}{MPT}              & \multicolumn{1}{c}{Phi-2}            & \multicolumn{1}{c|}{Falcon}          & \multicolumn{1}{c}{Llama2}          & \multicolumn{1}{c}{MPT}              & \multicolumn{1}{c}{Phi-2}            & \multicolumn{1}{c}{Falcon}          \\ \midrule
0          & \textbf{0.5642} & 0.2148           & \textbf{0.4438} & \textbf{0.4330} & \textbf{0.5642} & \textbf{0.2148} & \textbf{0.4438} & \textbf{0.4330} & \textbf{0.5642} & \textbf{0.2148} & \textbf{0.4438} & \textbf{0.4330} \\
1          & 0.4586          & 0.1976           & 0.3585          & 0.3469          & no-mid               & no-mid               & no-mid               & no-mid               & 0.4283          & 0.1791          & 0.4227          & 0.3602          \\
2          & 0.3455          & 0.1913           & 0.3430          & 0.3246          & 0.3322          & 0.1802          & 0.3375          & 0.2823          & 0.3974          & 0.2002          & 0.3975          & 0.3111          \\
4          & 0.2745          & \textbf{0.2209*} & 0.3019          & 0.2670          & 0.2857          & 0.1775          & 0.2885          & 0.2378          & 0.3795          & 0.2059*         & 0.3701          & 0.2736          \\
6          & 0.2898          & 0.2171*          & 0.2943          & 0.2392          & 0.2698          & 0.1424          & 0.2625          & 0.2103          & 0.3880          & 0.1892          & 0.3623          & 0.2656          \\
8          & 0.2643          & 0.2077*          & 0.2513          & 0.1878          & 0.2268          & 0.1002          & 0.2360          & 0.1745          & 0.3748          & 0.1944          & 0.3423          & 0.2424          \\
10         & 0.2537          & -                & -               & -               & 0.2180          & -               & -               & -               & 0.3716          & -               & -               & -               \\
12         & 0.2688          & -                & -               & -               & 0.2382          & -               & -               & -               & 0.3991          & -               & -               & -               \\
14         & 0.2583          & -                & -               & -               & 0.2280          & -               & -               & -               & 0.4118          & -               & -               & -               \\
16         & 0.2413          & -                & -               & -               & 0.2024          & -               & -               & -               & 0.3889          & -               & -               & -               \\
18         & 0.2348          & -                & -               & -               & 0.1795          & -               & -               & -               & 0.3781          & -               & -               & -               \\ \bottomrule
\end{tabular}
\end{table*}

\section{Results}
\label{sec:res_i}


\looseness -1 
Studying the characteristics of optimal prompts for RAG systems corresponds to answering our research question \textbf{(RQ)}: "\textit{What characteristics are desirable in a retriever to optimize prompt construction for RAG systems in order to increase the LLM effectiveness?"}. More specifically, we focus on three essential elements of the configuration: type, number, and positioning of the documents, and for each, we test various prompt combinations. 
%
To facilitate the understanding of our experimental setup, we employ a streamlined schema for representing the composition of prompts via the following symbols: 
\textbf{[I, \faStar, \faChain, \faChainBroken, \dice{6}, Q]}.
The task instruction (\textbf{I}) and the query (\textbf{Q}) are consistently positioned at the beginning and end, respectively. The middle section varies and represents different contextual elements - in this instance, these are gold, relevant, distracting, and random, appearing in that specific sequence. Additionally, the number of contextual documents is a variable in its own right and will be reported in the results tables below.

\subsection{Impact of Distracting Documents}
\label{sec:related}

\begin{figure}[htbp]
\begin{tcolorbox}[title=LLM Input - Distracting \faChainBroken \ and Gold \faStar]
\emph{Task Instruction...}\\
\textbf{Documents:}\\
\textbf{Document [1]}(Title: Han Solo) Before the events of the film, he and Chewbacca had lost the ``Millennium Falcon'' to thieves, but they reclaim the ship after it...\\
\textbf{Document [2]}(Title: Millennium Falcon) The ``Falcon'' has been depicted many times in the franchise, and ownership has changed several times...

\textcolor{gold}{\textbf{Document [3]}}(Title: Millennium Falcon) Han Solo won the Millennium Falcon from \textcolor{gold}{Lando Calrissian} in the card game sabacc...
\\ \\
\textbf{Question:} who owned the millennium falcon before han solo

\textbf{Answer:} \textcolor{red}{Han Solo}
\end{tcolorbox} 
\caption{Example LLM input with an erroneous output, highlighted in \textcolor{red}{red}. The context of the prompt is composed of distracting documents and the gold near the query. The task instruction is as in Figure \ref{fig:prompt_only_gold}.}

\label{fig:prompt_related_gold}
\end{figure}

\begin{figure}
    \centering
    \includegraphics[width=1\linewidth]{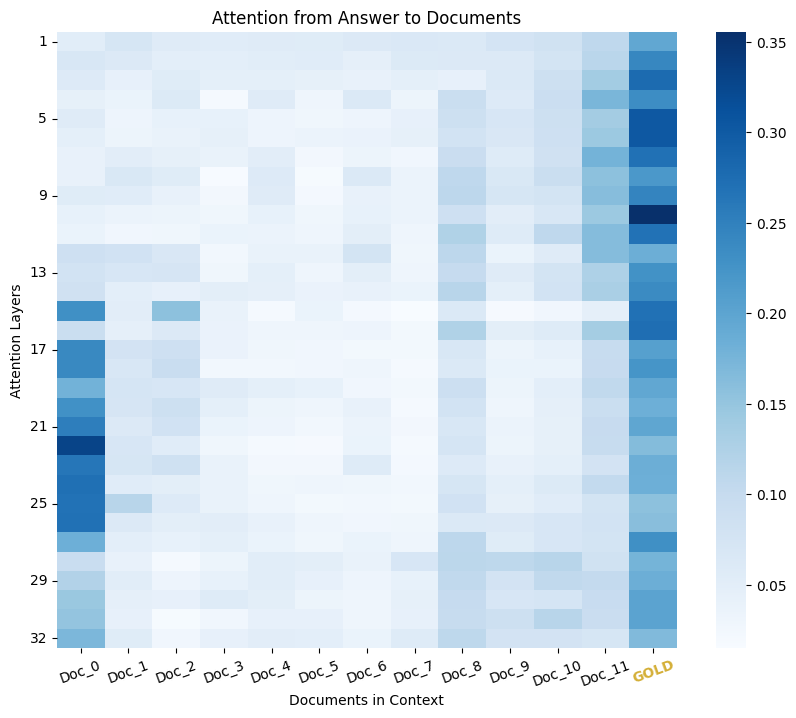}
    \caption{This heatmap depicts the attention distribution across the context documents from the example shown in Figure \ref{fig:prompt_related_gold}, relative to the answer generated by Llama2-7b in a prompt structured as [I, \faChainBroken, \faStar, Q]. Cell (i, j) denotes the mean attention that tokens in the generated answer allocate to the tokens of the i-th document within the j-th attention layer. This mean attention for each document is calculated by averaging the attention scores across all its constituent tokens.}
    \label{fig:attn}
\end{figure}

In our first set of experiments, we use a selection of 10K queries from the training set of the NQ-open dataset and assume an oracle setup in which the gold document for the query is known.
To this effect, we add to the gold document a set of distracting documents, i.e., documents with high retrieval scores but not containing the answer, in order to measure their impact on the system; schematically \textbf{[I, \faChainBroken, \faStar, Q]}.
Figure \ref{fig:prompt_related_gold} shows an example of this setup's visualization.
Results of this experiment are shown in Table \ref{tab:rel_table} (far, mid, and near relate to the distance between the gold document and the query; more details in the following sub-section).
A critical observation emerging from this analysis is a clear pattern of progressive accuracy degradation as the number of distracting documents included in the context increases. This was observed across all LLMs, with accuracy deteriorating by more than 0.38 ($-67\%$) in some cases.
Even more importantly, adding just one distracting document causes a sharp reduction in accuracy, with peaks of 0.24 ($-25\%$), as can be seen by comparing the row with $0$ distracting documents (only gold scenario, as seen in Figure \ref{fig:prompt_only_gold}) with that of $1$ distracting document.
This experiment highlights a critical issue for RAG systems, particularly in real-world IR settings where related but non-answer-containing documents are commonplace. Our empirical analysis suggests that introducing semantically aligned yet non-relevant documents adds a layer of complexity, potentially misguiding LLMs away from the correct response.
A visual explanation can be seen in Figure \ref{fig:attn}, which illustrates the attention scores within the prompt's context for a specific example in which the LLM incorrectly answers. This figure highlights the model's disproportionate focus on a distracting document (leftmost) at the expense of the gold document (rightmost), likely contributing to the erroneous response.
Note that for consistency of results across LLMs, we need to account for their various input token capabilities: Llama2 can process up to 4096 tokens, but other models are limited to 2048 tokens. This led to the exclusion of evaluations with a higher number of distracting documents (namely greater than 10)
as reflected by the empty values in the tables.

\begin{table*}
\caption{Accuracy results of the LLMs when evaluated with prompts composed of the gold document \faStar \ and a varying number of random \dice{6} documents. Surprisingly, increasing the number of random documents in the Near setting improves LLM's performance. Scenarios where the prompt exceeded the model's input limit, leading to potential data truncation, are not included ( - ). All values \textit{not} marked with an asterisk * denote statistically significant changes from the gold-only document scenario [I, \faStar, Q] (first row), as determined by a Wilcoxon test (p-value < 0.01). Additionally, the closed-book accuracy scores for the models are as follows: Llama2 (0.1123), MPT (0.1205), Phi-2 (0.0488), Falcon (0.1083).}
\label{tab:irr_table}
\begin{tabular}{r|cccc|cccc|cccc@{}}
\toprule
\multicolumn{1}{l|}{} & \multicolumn{4}{c|}{\textbf{Far - [I, \faStar, \dice{6}, Q]}}                                      & \multicolumn{4}{c|}{\textbf{Mid -  [I, \dice{6}, \faStar, \dice{6}, Q]}}                                     & \multicolumn{4}{c}{\textbf{Near -  [I, \dice{6}, \faStar, Q]}}                         \\ \midrule
\textbf{\# \dice{6}}             & Llama2          & MPT             & Phi-2             & Falcon          & Llama2          & MPT             & Phi-2             & Falcon          & Llama2          & MPT             & Phi-2             & Falcon          \\ \midrule
0                     & \textbf{0.5642} & 0.2148          & \textbf{0.4438} & \textbf{0.4330} & \textbf{0.5642} & 0.2148          & \textbf{0.4438} & \textbf{0.4330} & 0.5642          & 0.2148          & 0.4438          & \textbf{0.4330} \\
1                     & 0.4733          & 0.2447          & 0.4329          & 0.4035          & no-mid               & no-mid               & no-mid               & no-mid               & 0.4862          & 0.2125*          & 0.4587          & 0.4091          \\
2                     & 0.3776          & 0.2639          & 0.4249          & 0.3805          & 0.3928          & \textbf{0.2584} & 0.4293          & 0.3612          & 0.5032          & 0.2660          & \textbf{0.4614} & 0.3912          \\
4                     & 0.3109          & 0.2933          & 0.4091          & 0.3468          & 0.3998          & 0.2577          & 0.3985          & 0.3462          & 0.5221          & \textbf{0.2930} & 0.4311          & 0.3949          \\
6                     & 0.3547          & 0.3036          & 0.4130          & 0.3250          & 0.4138          & 0.2265          & 0.3891          & 0.3196          & 0.5681*         & 0.2890          & 0.4388          & 0.3908          \\
8                     & 0.3106          & \textbf{0.3039} & 0.3812          & 0.2543          & 0.3734          & 0.1566          & 0.3596          & 0.2767          & 0.5609*         & 0.2911          & 0.4258          & 0.3704          \\
10                    & 0.3390          & -               & -               & -               & 0.3675          & -               & -               & -               & 0.5579*         & -               & -               & -               \\
12                    & 0.3736          & -               & -               & -               & 0.3641          & -               & -               & -               & 0.5836            & -               & -               & -               \\
14                    & 0.3527          & -               & -               & -               & 0.3372          & -               & -               & -               & \textbf{0.5859} & -               & -               & -               \\
16                    & 0.3401          & -               & -               & -               & 0.3159          & -               & -               & -               & 0.5722          & -               & -               & -               \\
18                    & 0.3466          & -               & -               & -               & 0.2982          & -               & -               & -               & 0.5588*         & -               & -               & -               \\ \bottomrule
\end{tabular}
\end{table*}
 
In addition, we wanted to verify that our results were not overly dependent on the type of dense retrieval system we used. We wanted, in particular, to check whether another dense retriever specifically trained on ``hard negatives" would better distinguish between directly relevant and distracting documents, potentially leading to different results. 
To explore this hypothesis, we used ADORE \cite{zhan2021adore}, a state-of-the-art retriever trained with ``dynamic hard negatives'', to select the distracting documents.
In scenarios with 1, 2, and 4 distracting documents 
in the \textbf{[I, \faChainBroken, \faStar, Q]} setting with Llama2, we obtain an accuracy of 0.4068, 0.3815, and 0.3626, respectively. This is significantly lower than the baseline accuracy of 0.5642, where no distracting documents were included, and than the results obtained with Contriever in the same settings.
We conclude from this that distinguishing between relevant and distracting information is a hard problem that cannot be mitigated simply by changing the dense retrieval method at this stage. 



\subsection{Impact of Gold Positioning}

We conduct here another experiment where we systematically shift the position of the gold document within the context to study its impact on the model's effectiveness.
%
We define the positions of the gold document as follows:
\begin{itemize}
    \item \textbf{Near:} placed adjacent to the query in the prompt \textbf{[I, \faChainBroken, \faStar, Q]} (as in Figure \ref{fig:prompt_related_gold})
    \item \textbf{Mid:} inserted in the middle of the context \textbf{[I, \faChainBroken, \faStar, \faChainBroken, Q]} 
    \item \textbf{Far:} positioned as far as possible from the query in the context \textbf{[I, \faStar, \faChainBroken, Q]}
\end{itemize}
Results in these settings partially corroborate evidence from \cite{liu2023lost}.
The accuracy is higher when the gold document is near the query, lower when the gold document is furthest from it, and lowest when the gold document is placed in the middle of the context. For instance, Llama2, with 18 distracting documents, reaches an accuracy of 0.37, 0.23, and 0.17, respectively.
These results are consistent across all models tested in the setting with distracting documents.


\begin{table*}[!ht]
\caption{Accuracy of Llama2-7b in configurations involving random Wikipedia documents and retrieved documents [I, \dice{6}, \faFileTextO, Q]. Rows denote the number of random documents \dice{6} added, and columns show the quantity of retrieved documents \faFileTextO. The left section reports results using Contriever, and the right section using BM25. Scenarios where the prompt exceeded the model's input limit, leading to potential data truncation, are not included ( - ). Each value \emph{not} marked with an asterisk * represents a statistically significant change from the base case of retrieved documents only [I, \faFileTextO, Q] (first row), as determined by a Wilcoxon test (p-value < 0.01).}
\label{tab:rag_wild_plain}
\resizebox{\textwidth}{!}{
\begin{tabular}{r|ccccccc|ccccccc}
\toprule
\multicolumn{1}{l|}{} & \multicolumn{7}{c|}{\textbf{Contriever}}                                                                                    & \multicolumn{7}{c}{\textbf{BM25}}                                                                                           \\
\midrule
\diagbox{\# \dice{6}}{\# \faFileTextO}            & \multicolumn{1}{c}{1} & \multicolumn{1}{c}{2} & \multicolumn{1}{c}{3} & \multicolumn{1}{c}{4} & \multicolumn{1}{c}{5} & \multicolumn{1}{c}{8}                                   & \multicolumn{1}{c|}{10}                                   & \multicolumn{1}{c}{1} & \multicolumn{1}{c}{2} & \multicolumn{1}{c}{3} & \multicolumn{1}{c}{4} & \multicolumn{1}{c}{5} & \multicolumn{1}{c}{8} & \multicolumn{1}{c}{10}                                  \\
\midrule
0                     & 0.1620          & 0.1866          & 0.1876          & 0.1866          & 0.1921          & 0.2198          & 0.2108          & 0.2008          & 0.2208          & 0.2084          & 0.2028          & 0.2243          & 0.2492          & 0.2447          \\
1                     & 0.1308          & 0.1616          & 0.1717          & 0.1893*         & 0.1987*         & 0.2153*         & 0.2146*         & 0.1568          & 0.1963          & 0.1921          & 0.2115          & 0.2295*         & 0.2475*         & 0.2506*         \\
2                     & 0.1315          & 0.1644          & 0.1859*         & 0.2008          & 0.2174          & 0.2156*         & 0.2368          & 0.1644          & 0.1973          & 0.2080*         & 0.2281          & 0.2558          & 0.2495*         & 0.2596          \\
3                     & 0.1301          & 0.1727          & 0.2008          & 0.2316          & 0.2201          & 0.2198          & 0.2409          & 0.1568          & 0.2063          & 0.2160          & 0.2520          & 0.2579          & 0.2644          & 0.2707          \\
5                     & 0.1464          & 0.2056          & 0.2233          & 0.2240          & 0.2150          & \textbf{0.2451} & \textbf{0.2482} & 0.1772          & 0.2402          & 0.2437          & 0.2520          & 0.2554          & 0.2804          & \textbf{0.2866} \\
8                     & 0.1734          & 0.2066          & 0.2336          & 0.2375          & 0.2454          & 0.2416          & 0.2364          & 0.1994          & 0.2451          & 0.2579          & 0.2769          & 0.2817          & \textbf{0.2859} & 0.2777          \\
10                    & 0.1796          & 0.2174          & 0.2450          & 0.2502          & \textbf{0.2499} & 0.2420          & -               & 0.2108          & 0.2589          & 0.2734          & 0.2835          & \textbf{0.2935} & 0.2853          & -               \\
15                    & 0.2018          & 0.2354          & 0.2551          & \textbf{0.2530} & -               & -               & -               & 0.2243          & 0.2686          & 0.2790          & \textbf{0.2928} & -               & -               & -               \\
16                    & 0.2032          & \textbf{0.2471} & \textbf{0.2558} & -               & -               & -               & -               & 0.2323          & 0.2662          & \textbf{0.2838} & -               & -               & -               & -               \\
17                    & 0.2039          & 0.2426          & -               & -               & -               & -               & -               & \textbf{0.2326} & \textbf{0.2693} & -               & -               & -               & -               & -               \\
18                    & \textbf{0.2073} & -               & -               & -               & -               & -               & -               & 0.2309          & -               & -               & -               & -               & -               & -               \\ \bottomrule
\end{tabular}
}
\end{table*}

\begin{figure}[htbp]
\begin{tcolorbox}[title=LLM Input - Random \dice{6} and Gold \faStar]
\emph{Task instruction...}

\textbf{Documents:}

\textbf{Document [140]}(Title: Richard Yates (novelist)) For much of his life, Yates's work met almost universal critical acclaim, yet not one of his books sold over 12,000 copies in...

\textbf{Document [242]}(Title: Android version history) Code name Version number Initial release date API level Security patches (No codename ) 1.0 September 23...

\textcolor{gold}{\textbf{Document [3]}}(Title: Millennium Falcon) Han Solo won the Millennium Falcon from \textcolor{gold}{Lando Calrissian} in the card game sabacc...
\\ \\
\textbf{Question:} who owned the millennium falcon before han solo

\textbf{Answer:} \textcolor{correct}{Lando Calrissian}
\end{tcolorbox} 
\caption{Example LLM input with a correct output, highlighted in \textcolor{correct}{green}. The context of the prompt is composed of random documents and the gold near the query. The task instruction is as in Figure \ref{fig:prompt_only_gold}.}

\label{fig:prompt_random_gold}
\end{figure}

\subsection{Impact of Noise}
\label{sec:noise}
We devise an additional experimental setting aimed at evaluating the robustness of the RAG system against noise.
To this effect, we take the gold document and add to it a certain number of documents picked at random from the corpus; see an example in Figure \ref{fig:prompt_random_gold}. Against our expectations, the performance does not deteriorate in the presence of noise, as can be seen in Table \ref{tab:irr_table}. 
Instead, we observe an improvement in performance under the best-performing setting (near \textbf{[I, \dice{6}, \faStar, Q]}), with an improvement of 0.08 ($+36\%$) in the case of MPT.
Furthermore, we observe that different models exhibit distinct behaviors. 
Both Llama2 and Phi-2 showed improvements in this setting when the noise is introduced furthest from the query. 
However, when the noise is positioned in the far \textbf{[I, \faStar, \dice{6}, Q]} and mid \textbf{[I, \dice{6}, \faStar, \dice{6}, Q]} settings, these models exhibit a decline in performance.
Notably, this performance degradation is much less accentuated when compared to the earlier setting with distracting documents. 
This suggests that while Llama2 and Phi-2 can effectively handle noise far from the query, their ability to sift through irrelevant information diminishes as the noise is placed closer to it.
The MPT model presented a unique response; it showed an improvement in performance under all settings.
Standing out from the rest, the Falcon model did not exhibit an improvement in performance as observed in other models with the introduction of noise. 
Peculiarly enough, Falcon and Llama2 do not consistently exhibit a ``lost in the middle'' phenomenon, having in some instances better accuracy in the mid than far setting, for instance, in the case with $8$ noisy documents added.
%

\subsection{RAG in Practice}
To address our primary Research Question \textbf{(RQ)} about the characteristics of an effective RAG retriever, and following the results reported above, we now consider a more realistic scenario than an oracle setup. 
Namely, given a query, we retrieve a set of documents that can be either relevant or distracting.
We then add random documents to this set of retrieved ones, schematically: \textbf{[I, \dice{6}, \faFileTextO, Q]}.
For this second set of experiments, we use the test set of the NQ-open dataset.
Results for this experiment, using Llama2, can be seen on the left side of Table \ref{tab:rag_wild_plain}.
These results show that, regardless of the number of retrieved documents, adding random documents up until the context length is filled is almost always beneficial, with gains in terms of accuracy up to 0.07 ($+35\%$) in the case of 4 retrieved documents.

\begin{table*}[!ht]
\caption{Accuracy of Llama2-7b in configurations involving random documents and retrieved documents by Contriever [I, \dice{6}, \faFileTextO, Q]. Rows denote the number of random documents \dice{6} added, and columns show the quantity of retrieved documents \faFileTextO. The left section reports results with random documents from Reddit and the right section with nonsensical sentences made up of random
words. Scenarios where the prompt exceeded the model's input limit, leading to potential data truncation, are not included ( - ). Each value \emph{not} marked with an asterisk * represents a statistically significant change from the base case of retrieved documents only [I, \faFileTextO, Q] (first row), as determined by a Wilcoxon test (p-value < 0.01).}
\label{tab:reddit_random}
\resizebox{\textwidth}{!}{
\begin{tabular}{r|ccccccc|ccccccc}
\toprule
\multicolumn{1}{l|}{} & \multicolumn{7}{c|}{\textbf{Random from Reddit}}                                                                                                                                                   & \multicolumn{7}{c}{\textbf{Random Words}}                                                                                                                                       \\
\midrule
\diagbox{\# \dice{6}}{\# \faFileTextO}            & \multicolumn{1}{c}{1} & \multicolumn{1}{c}{2} & \multicolumn{1}{c}{3} & \multicolumn{1}{c}{4} & \multicolumn{1}{c}{5} & \multicolumn{1}{c}{8}                                   & \multicolumn{1}{c|}{10}                                   & \multicolumn{1}{c}{1} & \multicolumn{1}{c}{2} & \multicolumn{1}{c}{3} & \multicolumn{1}{c}{4} & \multicolumn{1}{c}{5} & \multicolumn{1}{c}{8} & \multicolumn{1}{c}{10}                                  \\
\midrule
0                     & 0.1620                & 0.1866                & 0.1876                & 0.1866                & 0.1921                & 0.2198          & 0.2108          & 0.1620                & 0.1866                & 0.1876                & 0.1866                & 0.1921                & 0.2198                & 0.2108          \\
1                     & 0.1693*               & 0.1931                & 0.1845*               & 0.1907                & 0.2008                & 0.2084          & 0.2084          & 0.1744                & 0.1924*                & 0.1969                & 0.2077                & 0.2091                & 0.2139*               & 0.2073*         \\
2                     & 0.1886                & 0.2018                & 0.2101                & 0.2143                & 0.2160                & 0.2222*                             & 0.2219          & 0.1765                & 0.1855*                & 0.2094                & 0.2122                & 0.2181                & 0.2045                & 0.2084*         \\
3                     & 0.1897                & 0.2108                & 0.2212                & 0.2340                & 0.2371                & 0.2326          & 0.2319          & 0.1755                & 0.1990                & 0.2166                & 0.2201                & 0.2288                & 0.2032                & 0.2156*         \\
5                     & 0.1897                & 0.2215                & 0.2388                & 0.2468                & 0.2409                & \textbf{0.2769} & \textbf{0.2451} & 0.1862                & 0.2139                & 0.2319                & 0.2367                & 0.2232                & 0.2184*               & 0.2278          \\
8                     & 0.2011                & 0.2326                & 0.2354                & 0.2489                & 0.2440                & 0.2568          & 0.2364                               & 0.1973                & 0.2274                & 0.2319                & 0.2316                & 0.2305                & 0.2357                & \textbf{0.2412} \\
10                    & 0.2053                & 0.2326                & 0.2451                & 0.2534                & \textbf{0.2551}       & 0.2658          & -                                    & 0.2053                & 0.2271                & 0.2340                & 0.2385                & \textbf{0.2406}       & \textbf{0.2499}       & -                                   \\
15                    & 0.2240                & 0.2489                & \textbf{0.2689}       & \textbf{0.2786}       & - & -                                   & -                                    & 0.2215                & 0.2416                & \textbf{0.2589}       & \textbf{0.2634}       & - & - & -                                   \\
16                    & 0.2240                & 0.2561                & 0.2676                & - & - & -                                   & -                                    & \textbf{0.2219}       & 0.2437                & 0.2568                & - & - & - & -                                   \\
17                    & \textbf{0.2243}       & \textbf{0.2565}       & - & - & - & -                                   & -                                    & 0.2201                & \textbf{0.2450}       & - & - & - & - & -                                   \\
18                    & 0.2240                & - & - & - & - & -                                   & -                                    & 0.2177                & - & - & - & - & - & -                                   \\
\bottomrule
\end{tabular}
}
\end{table*}

\subsubsection{Testing Sparse Retrievers}
In an effort to validate our initial observations, we replicate our experiment using a sparse retrieval approach, specifically BM25. The corresponding results are outlined in the right section of Table \ref{tab:rag_wild_plain}. Consistent with earlier findings, we observe that including random documents leads to an improvement in the effectiveness of the LLM. Notably, the use of BM25 yields an average increase in accuracy of 3-4 percentage points.
This improvement is attributed to the quality of documents retrieved by BM25. 
We quantitatively evaluate the effectiveness of the retrieval methods by computing the top-$k$ accuracy for varying numbers of retrieved documents.
Note that this heuristic, while indicative, does not capture the full spectrum of relevance. Our evaluation, based on the presence of correct answers within documents, might overlook the context-specific relevance due to potential lexical matches of the answer string in 
documents. Despite this limitation, this method aligns with established computational practices in literature \cite{karpukhin2020dense, izacard2021unsupervised}.
In our analysis, BM25 demonstrated higher relative top-$k$ accuracy (0.2966, 0.4105, 0.5237, 0.6663 for $k = 1, 2, 4, 10$) compared to those of Contriever (0.2502, 0.3569, 0.4784, 0.6085 for the same $k$), underscoring its effectiveness in retrieving more relevant documents in our experimental setup.

\subsubsection{Increasing The Randomness}

While our previous experiments show the benefits of adding random documents, one might argue that these documents are not totally random as they originate from the same corpus (Wikipedia) and that they might help the LLM answer in a fashion that is consistent with the corpus.
For this reason, we carry out another experiment in which random documents are drawn from a drastically different corpus in terms of tone and style, namely Reddit Webis-TLDR-17 dataset \cite{reddit}.
The results are outlined on the left of Table \ref{tab:reddit_random}.
The inclusion of documents from the Reddit corpus not only maintains the observed increase in accuracy but even enhances it, with an improvement of 0.023 ($+9\%$ accuracy) when compared to the previous best score.
Pushing the randomness even further, we carry out another test where we consider nonsensical sentences made up of random words as random documents.
Remarkably, even in this scenario, we observe a performance improvement when compared to the base case of Wikipedia random documents, as shown in the right side of Table \ref{tab:reddit_random}. 

\subsubsection{Falcon}


\begin{table}[!ht]
\caption{
Accuracy of Falcon-7b on Reddit data in the random + retrieved setting [I, \dice{6}, \faFileTextO, Q].
Rows denote the number of random documents \dice{6} added, and columns show the quantity of retrieved documents \faFileTextO. 
Scenarios where the prompt exceeded the model's input limit, leading to potential data truncation, are not included ( - ). 
Each value \emph{not} marked with an asterisk * represents a statistically significant change from the base case (first row), as determined by a Wilcoxon test (p-value < 0.05).
}
\label{tab:falcon}
\resizebox{\columnwidth}{!}{
\begin{tabular}{r|ccccccc}
\toprule
\diagbox{\# \dice{6}}{\# \faFileTextO} & 1      & 2               & 3               & 4               & 5      & 9               \\ \midrule
0 & 0.1568          & 0.1717          & 0.1855          & 0.1938          & 0.1942          & \textbf{0.1998} \\
1 & 0.1551*         & 0.1793*         & 0.1897*         & 0.1924*         & 0.1976*         & -               \\
2 & 0.1529*         & 0.1762*         & 0.1938*         & 0.2011*         & 0.1976*         & -               \\
3 & 0.1599*         & 0.1727*         & 0.1911*         & 0.2021*         & \textbf{0.2118} & -               \\
4 & 0.1606*         & 0.1758*         & 0.1959          & 0.2073          & 0.2108          & -               \\
5 & 0.1627*         & 0.1762*         & 0.2000          & \textbf{0.2108} & -               & -               \\
6 & 0.1651*         & 0.1848          & \textbf{0.2004} & -               & -               & -               \\
7 & 0.1675          & \textbf{0.1848} & -               & -               & -               & -               \\
8 & \textbf{0.1682} & -               & -               & -               & -               & -               \\ \bottomrule
\end{tabular}
}
\end{table}

As shown in Table \ref{tab:irr_table}, Falcon does not reach the same performance increase when random documents are added to the gold document \textbf{[I, \dice{6}, \faStar, Q]}. 
Accordingly, we want to verify whether it behaves differently when adding retrieved rather than gold documents.
We find that the addition of random documents on top of retrieved documents \textbf{[I, \dice{6}, \faFileTextO, Q]} does improve the effectiveness of Falcon; see detailed results in Table \ref{tab:falcon}.
These results are in contrast with the ones obtained in the oracle setting, where Falcon was robust to noise. 
This new finding further validates our experimental evidence, namely that, outside the oracle setting, all the tested models show an improvement when a certain amount of noise is added.

\subsection{Retriever Trade-Off}

The experimental evidence detailed above not only contradicts the common perception that semantically close documents are helpful for LLMs but also highlights the need for a delicate balance between relevant and random documents.
When arranged as described, random documents seem to exert a positive influence on LLM accuracy. 
However, for the LLM to generate accurate answers, some degree of relevant information must exist in the context.
On the other hand, an overabundance of retrieved documents increases the likelihood of including distracting and non-relevant information, leading to a sharp decline in performance.
While establishing a formal or comprehensive theory behind these findings remains an open research challenge, we can still infer that there seems to be a trade-off between the number of relevant and totally irrelevant documents. More specifically, we observed that the best effectiveness is achieved when a minimal set of documents is initially retrieved and then supplemented with random documents until the context limit is reached. For the queries examined in this study, retrieving between 3 and 5 documents is the most effective choice. Adding more increases the risk of including too many distracting, thus counterproductive, documents.
We argue here that there is a pressing need for further research towards investigating how these initial findings can be exploited. More importantly, it is evident that we have yet to refine our understanding of the retriever's role within a RAG system.

\subsubsection*{On The Unreasonable Effectiveness Of Random Documents}
We cannot close this paper without attempting to explain the results shown up to this point.
We refer back to our RAG formulation, particularly the conditioned function $p_\theta (y | \cdot, d)$.
In hindsight, we can now state that by adding random documents to the context, we are better conditioning this function, inducing enhanced accuracy.
Previous research \cite{attanasio2022entropy,hoffmann2023eureka}, particularly \cite{zhai2023stabilizing}, hints that there might be cases in which a pathologically low attention entropy causes the LLM to generate degenerate outputs with a sharp decrease in performance. These episodes are named entropy collapse.
Following this line of research, we measure the entropy of the attention scores in the case where only the gold document is supplied \textbf{[I, \faStar, Q]} against the case in which random documents are added \textbf{[I, \dice{6}, \faStar, Q]}.
We find that when we introduce random documents, the entropy of the systems has a 3X increase.
Although these experiments show a pattern, we cannot yet answer this question in a definitive manner.
While out of the scope of this work, which focuses on the retriever component of RAG systems, we believe it is highly important to investigate the reasons for which the LLM shows this behavior. 
Future studies should aim to elucidate why this noisy state is more advantageous and identify the characteristics that contribute to its effectiveness.

\section{Conclusions}

In this paper, we conducted the first comprehensive study focusing on the impact of retrieved documents on the RAG framework, aiming to understand the traits required in a retriever to optimize prompt construction for a RAG system.
This study led to several important findings, including two unexpected ones. First, the position of relevant information should be placed near the query; otherwise, the model seriously struggles to attend to it. Second, in contrast to common perception, top-scoring retrieved documents that do not contain the answer, when added to a prompt, negatively impact the LLM effectiveness. Finally, and even more surprisingly, random, noisy documents are actually helpful in increasing the accuracy of these systems when correctly positioned within a prompt.
While we have proposed heuristics to exploit these findings, further research is needed both to uncover the inner mechanisms behind this behavior and to develop a new generation of information retrieval techniques that are specifically designed to interact with the generative component.

\section*{Acknowledgments}
This work is supported by the Spoke ``FutureHPC \& BigData'' of the ICSC – Centro Nazionale di Ricerca in High-Performance Computing, Big Data and Quantum Computing, the Spoke ``Human-centered AI'' of the M4C2 - Investimento 1.3, Partenariato Esteso PE00000013 - "FAIR - Future Artificial Intelligence Research", SERICS (PE00000014), IR0000013 - SoBigData.it, funded by European Union – NextGenerationEU, the FoReLab project (Departments of Excellence), and the NEREO PRIN project funded by the Italian Ministry of Education and Research Grant no. 2022AEFHAZ. This work was carried out while Florin Cuconasu was enrolled in the Italian National Doctorate on Artificial Intelligence run by the Sapienza University of Rome.




\bibliographystyle{ACM-Reference-Format}
\bibliography{bib}


\begin{thebibliography}{60}


\ifx \showCODEN    \undefined \def \showCODEN     #1{\unskip}     \fi
\ifx \showDOI      \undefined \def \showDOI       #1{#1}\fi
\ifx \showISBNx    \undefined \def \showISBNx     #1{\unskip}     \fi
\ifx \showISBNxiii \undefined \def \showISBNxiii  #1{\unskip}     \fi
\ifx \showISSN     \undefined \def \showISSN      #1{\unskip}     \fi
\ifx \showLCCN     \undefined \def \showLCCN      #1{\unskip}     \fi
\ifx \shownote     \undefined \def \shownote      #1{#1}          \fi
\ifx \showarticletitle \undefined \def \showarticletitle #1{#1}   \fi
\ifx \showURL      \undefined \def \showURL       {\relax}        \fi
\providecommand\bibfield[2]{#2}
\providecommand\bibinfo[2]{#2}
\providecommand\natexlab[1]{#1}
\providecommand\showeprint[2][]{arXiv:#2}

\bibitem[Almazrouei et~al\mbox{.}(2023)]%
        {almazrouei2023falcon}
\bibfield{author}{\bibinfo{person}{Ebtesam Almazrouei}, \bibinfo{person}{Hamza Alobeidli}, \bibinfo{person}{Abdulaziz Alshamsi}, \bibinfo{person}{Alessandro Cappelli}, \bibinfo{person}{Ruxandra Cojocaru}, \bibinfo{person}{Mérouane Debbah}, \bibinfo{person}{Étienne Goffinet}, \bibinfo{person}{Daniel Hesslow}, \bibinfo{person}{Julien Launay}, \bibinfo{person}{Quentin Malartic}, \bibinfo{person}{Daniele Mazzotta}, \bibinfo{person}{Badreddine Noune}, \bibinfo{person}{Baptiste Pannier}, {and} \bibinfo{person}{Guilherme Penedo}.} \bibinfo{year}{2023}\natexlab{}.
\newblock \bibinfo{title}{The Falcon Series of Open Language Models}.
\newblock
\newblock
\showeprint[arxiv]{2311.16867}~[cs.CL]


\bibitem[Asai et~al\mbox{.}(2023)]%
        {asai2023selfrag}
\bibfield{author}{\bibinfo{person}{Akari Asai}, \bibinfo{person}{Zeqiu Wu}, \bibinfo{person}{Yizhong Wang}, \bibinfo{person}{Avirup Sil}, {and} \bibinfo{person}{Hannaneh Hajishirzi}.} \bibinfo{year}{2023}\natexlab{}.
\newblock \bibinfo{title}{Self-RAG: Learning to Retrieve, Generate, and Critique through Self-Reflection}.
\newblock
\newblock
\showeprint[arxiv]{2310.11511}~[cs.CL]


\bibitem[Attanasio et~al\mbox{.}(2022)]%
        {attanasio2022entropy}
\bibfield{author}{\bibinfo{person}{Giuseppe Attanasio}, \bibinfo{person}{Debora Nozza}, \bibinfo{person}{Dirk Hovy}, {and} \bibinfo{person}{Elena Baralis}.} \bibinfo{year}{2022}\natexlab{}.
\newblock \bibinfo{title}{Entropy-based attention regularization frees unintended bias mitigation from lists}.
\newblock
\newblock


\bibitem[Bacciu et~al\mbox{.}(2023a)]%
        {rraml}
\bibfield{author}{\bibinfo{person}{Andrea Bacciu}, \bibinfo{person}{Florin Cuconasu}, \bibinfo{person}{Federico Siciliano}, \bibinfo{person}{Fabrizio Silvestri}, \bibinfo{person}{Nicola Tonellotto}, {and} \bibinfo{person}{Giovanni Trappolini}.} \bibinfo{year}{2023}\natexlab{a}.
\newblock \showarticletitle{{RRAML:} Reinforced Retrieval Augmented Machine Learning}. In \bibinfo{booktitle}{\emph{Proceedings of the Discussion Papers - 22nd International Conference of the Italian Association for Artificial Intelligence (AIxIA 2023 {DP)} co-located with 22nd International Conference of the Italian Association for Artificial Intelligence (AIxIA 2023), Rome, Italy, November 6-9, 2023}} \emph{(\bibinfo{series}{{CEUR} Workshop Proceedings}, Vol.~\bibinfo{volume}{3537})}, \bibfield{editor}{\bibinfo{person}{Roberto Basili}, \bibinfo{person}{Domenico Lembo}, \bibinfo{person}{Carla Limongelli}, {and} \bibinfo{person}{Andrea Orlandini}} (Eds.). \bibinfo{publisher}{CEUR-WS.org}, \bibinfo{pages}{29--37}.
\newblock
\urldef\tempurl%
\url{https://ceur-ws.org/Vol-3537/paper4.pdf}
\showURL{%
\tempurl}


\bibitem[Bacciu et~al\mbox{.}(2023b)]%
        {fauno}
\bibfield{author}{\bibinfo{person}{Andrea Bacciu}, \bibinfo{person}{Giovanni Trappolini}, \bibinfo{person}{Andrea Santilli}, \bibinfo{person}{Emanuele Rodol{\`{a}}}, {and} \bibinfo{person}{Fabrizio Silvestri}.} \bibinfo{year}{2023}\natexlab{b}.
\newblock \showarticletitle{Fauno: The Italian Large Language Model that will leave you senza parole!}. In \bibinfo{booktitle}{\emph{Proceedings of the 13th Italian Information Retrieval Workshop {(IIR} 2023), Pisa, Italy, June 8-9, 2023}} \emph{(\bibinfo{series}{{CEUR} Workshop Proceedings}, Vol.~\bibinfo{volume}{3448})}, \bibfield{editor}{\bibinfo{person}{Franco~Maria Nardini}, \bibinfo{person}{Nicola Tonellotto}, \bibinfo{person}{Guglielmo Faggioli}, {and} \bibinfo{person}{Antonio Ferrara}} (Eds.). \bibinfo{publisher}{CEUR-WS.org}, \bibinfo{pages}{9--17}.
\newblock
\urldef\tempurl%
\url{https://ceur-ws.org/Vol-3448/paper-24.pdf}
\showURL{%
\tempurl}


\bibitem[Beeching et~al\mbox{.}(2023)]%
        {open-llm-leaderboard}
\bibfield{author}{\bibinfo{person}{Edward Beeching}, \bibinfo{person}{Clémentine Fourrier}, \bibinfo{person}{Nathan Habib}, \bibinfo{person}{Sheon Han}, \bibinfo{person}{Nathan Lambert}, \bibinfo{person}{Nazneen Rajani}, \bibinfo{person}{Omar Sanseviero}, \bibinfo{person}{Lewis Tunstall}, {and} \bibinfo{person}{Thomas Wolf}.} \bibinfo{year}{2023}\natexlab{}.
\newblock \bibinfo{title}{Open LLM Leaderboard}.
\newblock \bibinfo{howpublished}{\url{https://huggingface.co/spaces/HuggingFaceH4/open_llm_leaderboard}}.
\newblock


\bibitem[BehnamGhader et~al\mbox{.}(2023)]%
        {behnamghader2022can}
\bibfield{author}{\bibinfo{person}{Parishad BehnamGhader}, \bibinfo{person}{Santiago Miret}, {and} \bibinfo{person}{Siva Reddy}.} \bibinfo{year}{2023}\natexlab{}.
\newblock \showarticletitle{Can Retriever-Augmented Language Models Reason? The Blame Game Between the Retriever and the Language Model}. In \bibinfo{booktitle}{\emph{Findings of the Association for Computational Linguistics: EMNLP 2023}}, \bibfield{editor}{\bibinfo{person}{Houda Bouamor}, \bibinfo{person}{Juan Pino}, {and} \bibinfo{person}{Kalika Bali}} (Eds.). \bibinfo{publisher}{Association for Computational Linguistics}, \bibinfo{address}{Singapore}, \bibinfo{pages}{15492--15509}.
\newblock
\urldef\tempurl%
\url{https://doi.org/10.18653/v1/2023.findings-emnlp.1036}
\showDOI{\tempurl}


\bibitem[Borgeaud et~al\mbox{.}(2022)]%
        {borgeaud2022improving}
\bibfield{author}{\bibinfo{person}{Sebastian Borgeaud}, \bibinfo{person}{Arthur Mensch}, \bibinfo{person}{Jordan Hoffmann}, \bibinfo{person}{Trevor Cai}, \bibinfo{person}{Eliza Rutherford}, \bibinfo{person}{Katie Millican}, \bibinfo{person}{George~Bm Van Den~Driessche}, \bibinfo{person}{Jean-Baptiste Lespiau}, \bibinfo{person}{Bogdan Damoc}, \bibinfo{person}{Aidan Clark}, {et~al\mbox{.}}} \bibinfo{year}{2022}\natexlab{}.
\newblock \showarticletitle{Improving language models by retrieving from trillions of tokens}. In \bibinfo{booktitle}{\emph{International conference on machine learning}}. \bibinfo{publisher}{PMLR}, \bibinfo{address}{Baltimora}, \bibinfo{pages}{2206--2240}.
\newblock


\bibitem[Brown et~al\mbox{.}(2020)]%
        {brown2020language}
\bibfield{author}{\bibinfo{person}{Tom Brown}, \bibinfo{person}{Benjamin Mann}, \bibinfo{person}{Nick Ryder}, \bibinfo{person}{Melanie Subbiah}, \bibinfo{person}{Jared~D Kaplan}, \bibinfo{person}{Prafulla Dhariwal}, \bibinfo{person}{Arvind Neelakantan}, \bibinfo{person}{Pranav Shyam}, \bibinfo{person}{Girish Sastry}, \bibinfo{person}{Amanda Askell}, {et~al\mbox{.}}} \bibinfo{year}{2020}\natexlab{}.
\newblock \showarticletitle{Language models are few-shot learners}.
\newblock \bibinfo{journal}{\emph{Advances in neural information processing systems}}  \bibinfo{volume}{33} (\bibinfo{year}{2020}), \bibinfo{pages}{1877--1901}.
\newblock


\bibitem[Cui et~al\mbox{.}(2023)]%
        {Chinese-LLaMA-Alpaca}
\bibfield{author}{\bibinfo{person}{Yiming Cui}, \bibinfo{person}{Ziqing Yang}, {and} \bibinfo{person}{Xin Yao}.} \bibinfo{year}{2023}\natexlab{}.
\newblock \showarticletitle{Efficient and Effective Text Encoding for Chinese LLaMA and Alpaca}.
\newblock \bibinfo{journal}{\emph{arXiv preprint arXiv:2304.08177}} (\bibinfo{year}{2023}).
\newblock
\urldef\tempurl%
\url{https://arxiv.org/abs/2304.08177}
\showURL{%
\tempurl}


\bibitem[Douze et~al\mbox{.}(2024)]%
        {douze2024faiss}
\bibfield{author}{\bibinfo{person}{Matthijs Douze}, \bibinfo{person}{Alexandr Guzhva}, \bibinfo{person}{Chengqi Deng}, \bibinfo{person}{Jeff Johnson}, \bibinfo{person}{Gergely Szilvasy}, \bibinfo{person}{Pierre-Emmanuel Mazaré}, \bibinfo{person}{Maria Lomeli}, \bibinfo{person}{Lucas Hosseini}, {and} \bibinfo{person}{Hervé Jégou}.} \bibinfo{year}{2024}\natexlab{}.
\newblock \showarticletitle{The Faiss library}.
\newblock  (\bibinfo{year}{2024}).
\newblock
\showeprint[arxiv]{2401.08281}~[cs.LG]


\bibitem[Garrachonr(2023)]%
        {Garrachonr_2023}
\bibfield{author}{\bibinfo{person}{Garrachonr}.} \bibinfo{year}{2023}\natexlab{}.
\newblock \bibinfo{title}{LlamaDos}.
\newblock \bibinfo{howpublished}{\url{https://github.com/Garrachonr/LlamaDos}}.
\newblock


\bibitem[Guu et~al\mbox{.}(2020)]%
        {guu2020retrieval}
\bibfield{author}{\bibinfo{person}{Kelvin Guu}, \bibinfo{person}{Kenton Lee}, \bibinfo{person}{Zora Tung}, \bibinfo{person}{Panupong Pasupat}, {and} \bibinfo{person}{Mingwei Chang}.} \bibinfo{year}{2020}\natexlab{}.
\newblock \showarticletitle{Retrieval augmented language model pre-training}. In \bibinfo{booktitle}{\emph{International conference on machine learning}}. \bibinfo{publisher}{PMLR}, \bibinfo{address}{Vienna}, \bibinfo{pages}{3929--3938}.
\newblock


\bibitem[Hoffmann et~al\mbox{.}(2023)]%
        {hoffmann2023eureka}
\bibfield{author}{\bibinfo{person}{David~T Hoffmann}, \bibinfo{person}{Simon Schrodi}, \bibinfo{person}{Nadine Behrmann}, \bibinfo{person}{Volker Fischer}, {and} \bibinfo{person}{Thomas Brox}.} \bibinfo{year}{2023}\natexlab{}.
\newblock \bibinfo{title}{Eureka-Moments in Transformers: Multi-Step Tasks Reveal Softmax Induced Optimization Problems}.
\newblock
\newblock


\bibitem[Izacard et~al\mbox{.}(2021)]%
        {izacard2021unsupervised}
\bibfield{author}{\bibinfo{person}{Gautier Izacard}, \bibinfo{person}{Mathilde Caron}, \bibinfo{person}{Lucas Hosseini}, \bibinfo{person}{Sebastian Riedel}, \bibinfo{person}{Piotr Bojanowski}, \bibinfo{person}{Armand Joulin}, {and} \bibinfo{person}{Edouard Grave}.} \bibinfo{year}{2021}\natexlab{}.
\newblock \bibinfo{title}{Unsupervised dense information retrieval with contrastive learning}.
\newblock
\newblock


\bibitem[Javaheripi et~al\mbox{.}(2023)]%
        {javaheripi2023phi}
\bibfield{author}{\bibinfo{person}{Mojan Javaheripi}, \bibinfo{person}{S{\'e}bastien Bubeck}, \bibinfo{person}{Marah Abdin}, \bibinfo{person}{Jyoti Aneja}, \bibinfo{person}{Sebastien Bubeck}, \bibinfo{person}{Caio C{\'e}sar~Teodoro Mendes}, \bibinfo{person}{Weizhu Chen}, \bibinfo{person}{Allie Del~Giorno}, \bibinfo{person}{Ronen Eldan}, \bibinfo{person}{Sivakanth Gopi}, {et~al\mbox{.}}} \bibinfo{year}{2023}\natexlab{}.
\newblock \bibinfo{title}{Phi-2: The surprising power of small language models}.
\newblock
\newblock


\bibitem[jphme(2023)]%
        {jphme_2023}
\bibfield{author}{\bibinfo{person}{jphme}.} \bibinfo{year}{2023}\natexlab{}.
\newblock \bibinfo{title}{Llama-2-13b-chat-german}.
\newblock \bibinfo{howpublished}{\url{https://huggingface.co/jphme/Llama-2-13b-chat-german}}.
\newblock


\bibitem[Kandpal et~al\mbox{.}(2023)]%
        {same-accuracy}
\bibfield{author}{\bibinfo{person}{Nikhil Kandpal}, \bibinfo{person}{Haikang Deng}, \bibinfo{person}{Adam Roberts}, \bibinfo{person}{Eric Wallace}, {and} \bibinfo{person}{Colin Raffel}.} \bibinfo{year}{2023}\natexlab{}.
\newblock \showarticletitle{Large language models struggle to learn long-tail knowledge}. In \bibinfo{booktitle}{\emph{Proceedings of the 40th International Conference on Machine Learning}} \emph{(\bibinfo{series}{ICML'23})}. \bibinfo{publisher}{JMLR.org}, \bibinfo{address}{Honolulu, Hawaii, USA}, Article \bibinfo{articleno}{641}, \bibinfo{numpages}{12}~pages.
\newblock


\bibitem[Karpukhin et~al\mbox{.}(2020a)]%
        {karpukhin2020dense}
\bibfield{author}{\bibinfo{person}{Vladimir Karpukhin}, \bibinfo{person}{Barlas O{\u{g}}uz}, \bibinfo{person}{Sewon Min}, \bibinfo{person}{Patrick Lewis}, \bibinfo{person}{Ledell Wu}, \bibinfo{person}{Sergey Edunov}, \bibinfo{person}{Danqi Chen}, {and} \bibinfo{person}{Wen-tau Yih}.} \bibinfo{year}{2020}\natexlab{a}.
\newblock \bibinfo{title}{Dense passage retrieval for open-domain question answering}.
\newblock
\newblock


\bibitem[Karpukhin et~al\mbox{.}(2020b)]%
        {DPR}
\bibfield{author}{\bibinfo{person}{Vladimir Karpukhin}, \bibinfo{person}{Barlas Oguz}, \bibinfo{person}{Sewon Min}, \bibinfo{person}{Patrick Lewis}, \bibinfo{person}{Ledell Wu}, \bibinfo{person}{Sergey Edunov}, \bibinfo{person}{Danqi Chen}, {and} \bibinfo{person}{Wen-tau Yih}.} \bibinfo{year}{2020}\natexlab{b}.
\newblock \showarticletitle{Dense Passage Retrieval for Open-Domain Question Answering}. In \bibinfo{booktitle}{\emph{Proceedings of the 2020 Conference on Empirical Methods in Natural Language Processing (EMNLP)}}, \bibfield{editor}{\bibinfo{person}{Bonnie Webber}, \bibinfo{person}{Trevor Cohn}, \bibinfo{person}{Yulan He}, {and} \bibinfo{person}{Yang Liu}} (Eds.). \bibinfo{publisher}{Association for Computational Linguistics}, \bibinfo{address}{Online}, \bibinfo{pages}{6769--6781}.
\newblock
\urldef\tempurl%
\url{https://doi.org/10.18653/v1/2020.emnlp-main.550}
\showDOI{\tempurl}


\bibitem[Ke et~al\mbox{.}(2024)]%
        {ke2024bridging}
\bibfield{author}{\bibinfo{person}{Zixuan Ke}, \bibinfo{person}{Weize Kong}, \bibinfo{person}{Cheng Li}, \bibinfo{person}{Mingyang Zhang}, \bibinfo{person}{Qiaozhu Mei}, {and} \bibinfo{person}{Michael Bendersky}.} \bibinfo{year}{2024}\natexlab{}.
\newblock \showarticletitle{Bridging the Preference Gap between Retrievers and LLMs}.
\newblock \bibinfo{journal}{\emph{arXiv preprint arXiv:2401.06954}} (\bibinfo{year}{2024}).
\newblock


\bibitem[Kenton and Toutanova(2019)]%
        {kenton2019bert}
\bibfield{author}{\bibinfo{person}{Jacob Devlin Ming-Wei~Chang Kenton} {and} \bibinfo{person}{Lee~Kristina Toutanova}.} \bibinfo{year}{2019}\natexlab{}.
\newblock \showarticletitle{BERT: Pre-training of deep bidirectional transformers for language understanding}. In \bibinfo{booktitle}{\emph{Proceedings of naacL-HLT}}, Vol.~\bibinfo{volume}{1}. \bibinfo{publisher}{Association for Computational Linguistic}, \bibinfo{address}{Minneapolis}, \bibinfo{pages}{2}.
\newblock


\bibitem[Khandelwal et~al\mbox{.}(2018)]%
        {khandelwal2018sharp}
\bibfield{author}{\bibinfo{person}{Urvashi Khandelwal}, \bibinfo{person}{He He}, \bibinfo{person}{Peng Qi}, {and} \bibinfo{person}{Dan Jurafsky}.} \bibinfo{year}{2018}\natexlab{}.
\newblock \bibinfo{title}{Sharp nearby, fuzzy far away: How neural language models use context}.
\newblock
\newblock


\bibitem[Khattab and Zaharia(2020)]%
        {khattab2020colbert}
\bibfield{author}{\bibinfo{person}{Omar Khattab} {and} \bibinfo{person}{Matei Zaharia}.} \bibinfo{year}{2020}\natexlab{}.
\newblock \showarticletitle{Colbert: Efficient and effective passage search via contextualized late interaction over bert}. In \bibinfo{booktitle}{\emph{Proceedings of the 43rd International ACM SIGIR conference on research and development in Information Retrieval}}. \bibinfo{publisher}{ACM}, \bibinfo{address}{Xi'an}, \bibinfo{pages}{39--48}.
\newblock


\bibitem[Koopman and Zuccon(2023)]%
        {koopman-zuccon-2023-dr}
\bibfield{author}{\bibinfo{person}{Bevan Koopman} {and} \bibinfo{person}{Guido Zuccon}.} \bibinfo{year}{2023}\natexlab{}.
\newblock \showarticletitle{Dr {C}hat{GPT} tell me what {I} want to hear: How different prompts impact health answer correctness}. In \bibinfo{booktitle}{\emph{Proceedings of the 2023 Conference on Empirical Methods in Natural Language Processing}}, \bibfield{editor}{\bibinfo{person}{Houda Bouamor}, \bibinfo{person}{Juan Pino}, {and} \bibinfo{person}{Kalika Bali}} (Eds.). \bibinfo{publisher}{Association for Computational Linguistics}, \bibinfo{address}{Singapore}, \bibinfo{pages}{15012--15022}.
\newblock
\urldef\tempurl%
\url{https://doi.org/10.18653/v1/2023.emnlp-main.928}
\showDOI{\tempurl}


\bibitem[Kwiatkowski et~al\mbox{.}(2019)]%
        {nq}
\bibfield{author}{\bibinfo{person}{Tom Kwiatkowski}, \bibinfo{person}{Jennimaria Palomaki}, \bibinfo{person}{Olivia Redfield}, \bibinfo{person}{Michael Collins}, \bibinfo{person}{Ankur Parikh}, \bibinfo{person}{Chris Alberti}, \bibinfo{person}{Danielle Epstein}, \bibinfo{person}{Illia Polosukhin}, \bibinfo{person}{Jacob Devlin}, \bibinfo{person}{Kenton Lee}, \bibinfo{person}{Kristina Toutanova}, \bibinfo{person}{Llion Jones}, \bibinfo{person}{Matthew Kelcey}, \bibinfo{person}{Ming-Wei Chang}, \bibinfo{person}{Andrew~M. Dai}, \bibinfo{person}{Jakob Uszkoreit}, \bibinfo{person}{Quoc Le}, {and} \bibinfo{person}{Slav Petrov}.} \bibinfo{year}{2019}\natexlab{}.
\newblock \showarticletitle{Natural Questions: A Benchmark for Question Answering Research}.
\newblock \bibinfo{journal}{\emph{Transactions of the Association for Computational Linguistics}}  \bibinfo{volume}{7} (\bibinfo{year}{2019}), \bibinfo{pages}{452--466}.
\newblock
\urldef\tempurl%
\url{https://doi.org/10.1162/tacl_a_00276}
\showDOI{\tempurl}


\bibitem[Lee et~al\mbox{.}(2019)]%
        {nq-open}
\bibfield{author}{\bibinfo{person}{Kenton Lee}, \bibinfo{person}{Ming{-}Wei Chang}, {and} \bibinfo{person}{Kristina Toutanova}.} \bibinfo{year}{2019}\natexlab{}.
\newblock \showarticletitle{Latent Retrieval for Weakly Supervised Open Domain Question Answering}. In \bibinfo{booktitle}{\emph{Proceedings of the 57th Conference of the Association for Computational Linguistics, {ACL} 2019, Florence, Italy, July 28- August 2, 2019, Volume 1: Long Papers}}, \bibfield{editor}{\bibinfo{person}{Anna Korhonen}, \bibinfo{person}{David~R. Traum}, {and} \bibinfo{person}{Llu{\'{\i}}s M{\`{a}}rquez}} (Eds.). \bibinfo{publisher}{Association for Computational Linguistics}, \bibinfo{address}{Florence}, \bibinfo{pages}{6086--6096}.
\newblock
\urldef\tempurl%
\url{https://doi.org/10.18653/V1/P19-1612}
\showDOI{\tempurl}


\bibitem[Lewis et~al\mbox{.}(2020)]%
        {lewis2020retrieval}
\bibfield{author}{\bibinfo{person}{Patrick Lewis}, \bibinfo{person}{Ethan Perez}, \bibinfo{person}{Aleksandra Piktus}, \bibinfo{person}{Fabio Petroni}, \bibinfo{person}{Vladimir Karpukhin}, \bibinfo{person}{Naman Goyal}, \bibinfo{person}{Heinrich K{\"u}ttler}, \bibinfo{person}{Mike Lewis}, \bibinfo{person}{Wen-tau Yih}, \bibinfo{person}{Tim Rockt{\"a}schel}, {et~al\mbox{.}}} \bibinfo{year}{2020}\natexlab{}.
\newblock \showarticletitle{Retrieval-augmented generation for knowledge-intensive nlp tasks}.
\newblock \bibinfo{journal}{\emph{Advances in Neural Information Processing Systems}}  \bibinfo{volume}{33} (\bibinfo{year}{2020}), \bibinfo{pages}{9459--9474}.
\newblock


\bibitem[Li et~al\mbox{.}(2023)]%
        {li2023textbooks}
\bibfield{author}{\bibinfo{person}{Yuanzhi Li}, \bibinfo{person}{S{\'e}bastien Bubeck}, \bibinfo{person}{Ronen Eldan}, \bibinfo{person}{Allie Del~Giorno}, \bibinfo{person}{Suriya Gunasekar}, {and} \bibinfo{person}{Yin~Tat Lee}.} \bibinfo{year}{2023}\natexlab{}.
\newblock \bibinfo{title}{Textbooks are all you need ii: phi-1.5 technical report}.
\newblock
\newblock


\bibitem[Liu et~al\mbox{.}(2023)]%
        {liu2023lost}
\bibfield{author}{\bibinfo{person}{Nelson~F Liu}, \bibinfo{person}{Kevin Lin}, \bibinfo{person}{John Hewitt}, \bibinfo{person}{Ashwin Paranjape}, \bibinfo{person}{Michele Bevilacqua}, \bibinfo{person}{Fabio Petroni}, {and} \bibinfo{person}{Percy Liang}.} \bibinfo{year}{2023}\natexlab{}.
\newblock \bibinfo{title}{Lost in the middle: How language models use long contexts}.
\newblock
\newblock


\bibitem[Lu et~al\mbox{.}(2022)]%
        {lu-etal-2022-fantastically}
\bibfield{author}{\bibinfo{person}{Yao Lu}, \bibinfo{person}{Max Bartolo}, \bibinfo{person}{Alastair Moore}, \bibinfo{person}{Sebastian Riedel}, {and} \bibinfo{person}{Pontus Stenetorp}.} \bibinfo{year}{2022}\natexlab{}.
\newblock \showarticletitle{Fantastically Ordered Prompts and Where to Find Them: Overcoming Few-Shot Prompt Order Sensitivity}. In \bibinfo{booktitle}{\emph{Proceedings of the 60th Annual Meeting of the Association for Computational Linguistics (Volume 1: Long Papers)}}, \bibfield{editor}{\bibinfo{person}{Smaranda Muresan}, \bibinfo{person}{Preslav Nakov}, {and} \bibinfo{person}{Aline Villavicencio}} (Eds.). \bibinfo{publisher}{Association for Computational Linguistics}, \bibinfo{address}{Dublin, Ireland}, \bibinfo{pages}{8086--8098}.
\newblock
\urldef\tempurl%
\url{https://doi.org/10.18653/v1/2022.acl-long.556}
\showDOI{\tempurl}


\bibitem[Manning et~al\mbox{.}(2008)]%
        {manning2008term}
\bibfield{author}{\bibinfo{person}{C Manning}, \bibinfo{person}{P Raghavan}, {and} \bibinfo{person}{H Schutze}.} \bibinfo{year}{2008}\natexlab{}.
\newblock \bibinfo{booktitle}{\emph{Term weighting, and the vector space model}}.
\newblock \bibinfo{publisher}{Cambridge University Press Cambridge}, \bibinfo{address}{Cambridge}. 109--133 pages.
\newblock


\bibitem[Mialon et~al\mbox{.}(2023)]%
        {mialon2023augmented}
\bibfield{author}{\bibinfo{person}{Gr{\'e}goire Mialon}, \bibinfo{person}{Roberto Dess{\`\i}}, \bibinfo{person}{Maria Lomeli}, \bibinfo{person}{Christoforos Nalmpantis}, \bibinfo{person}{Ram Pasunuru}, \bibinfo{person}{Roberta Raileanu}, \bibinfo{person}{Baptiste Rozi{\`e}re}, \bibinfo{person}{Timo Schick}, \bibinfo{person}{Jane Dwivedi-Yu}, \bibinfo{person}{Asli Celikyilmaz}, {et~al\mbox{.}}} \bibinfo{year}{2023}\natexlab{}.
\newblock \bibinfo{title}{Augmented language models: a survey}.
\newblock
\newblock


\bibitem[Min et~al\mbox{.}(2020)]%
        {ambigqa}
\bibfield{author}{\bibinfo{person}{Sewon Min}, \bibinfo{person}{Julian Michael}, \bibinfo{person}{Hannaneh Hajishirzi}, {and} \bibinfo{person}{Luke Zettlemoyer}.} \bibinfo{year}{2020}\natexlab{}.
\newblock \showarticletitle{{A}mbig{QA}: Answering Ambiguous Open-domain Questions}. In \bibinfo{booktitle}{\emph{Proceedings of the 2020 Conference on Empirical Methods in Natural Language Processing (EMNLP)}}, \bibfield{editor}{\bibinfo{person}{Bonnie Webber}, \bibinfo{person}{Trevor Cohn}, \bibinfo{person}{Yulan He}, {and} \bibinfo{person}{Yang Liu}} (Eds.). \bibinfo{publisher}{Association for Computational Linguistics}, \bibinfo{address}{Online}, \bibinfo{pages}{5783--5797}.
\newblock
\urldef\tempurl%
\url{https://doi.org/10.18653/v1/2020.emnlp-main.466}
\showDOI{\tempurl}


\bibitem[Penedo et~al\mbox{.}(2023)]%
        {penedo2023refinedweb}
\bibfield{author}{\bibinfo{person}{Guilherme Penedo}, \bibinfo{person}{Quentin Malartic}, \bibinfo{person}{Daniel Hesslow}, \bibinfo{person}{Ruxandra Cojocaru}, \bibinfo{person}{Alessandro Cappelli}, \bibinfo{person}{Hamza Alobeidli}, \bibinfo{person}{Baptiste Pannier}, \bibinfo{person}{Ebtesam Almazrouei}, {and} \bibinfo{person}{Julien Launay}.} \bibinfo{year}{2023}\natexlab{}.
\newblock \bibinfo{title}{The RefinedWeb Dataset for Falcon LLM: Outperforming Curated Corpora with Web Data, and Web Data Only}.
\newblock
\newblock
\showeprint[arxiv]{2306.01116}~[cs.CL]


\bibitem[Press et~al\mbox{.}(2022)]%
        {press2022alibi}
\bibfield{author}{\bibinfo{person}{Ofir Press}, \bibinfo{person}{Noah Smith}, {and} \bibinfo{person}{Mike Lewis}.} \bibinfo{year}{2022}\natexlab{}.
\newblock \bibinfo{title}{Train Short, Test Long: Attention with Linear Biases Enables Input Length Extrapolation}.
\newblock
\newblock
\urldef\tempurl%
\url{https://openreview.net/forum?id=R8sQPpGCv0}
\showURL{%
\tempurl}


\bibitem[Radford et~al\mbox{.}(2018)]%
        {radford2018improving}
\bibfield{author}{\bibinfo{person}{Alec Radford}, \bibinfo{person}{Karthik Narasimhan}, \bibinfo{person}{Tim Salimans}, \bibinfo{person}{Ilya Sutskever}, {et~al\mbox{.}}} \bibinfo{year}{2018}\natexlab{}.
\newblock \bibinfo{title}{Improving language understanding by generative pre-training}.
\newblock
\newblock


\bibitem[Radford et~al\mbox{.}(2019)]%
        {radford2019language}
\bibfield{author}{\bibinfo{person}{Alec Radford}, \bibinfo{person}{Jeffrey Wu}, \bibinfo{person}{Rewon Child}, \bibinfo{person}{David Luan}, \bibinfo{person}{Dario Amodei}, \bibinfo{person}{Ilya Sutskever}, {et~al\mbox{.}}} \bibinfo{year}{2019}\natexlab{}.
\newblock \showarticletitle{Language models are unsupervised multitask learners}.
\newblock \bibinfo{journal}{\emph{OpenAI blog}} \bibinfo{volume}{1}, \bibinfo{number}{8} (\bibinfo{year}{2019}), \bibinfo{pages}{9}.
\newblock


\bibitem[Ram et~al\mbox{.}(2023)]%
        {ram2023context}
\bibfield{author}{\bibinfo{person}{Ori Ram}, \bibinfo{person}{Yoav Levine}, \bibinfo{person}{Itay Dalmedigos}, \bibinfo{person}{Dor Muhlgay}, \bibinfo{person}{Amnon Shashua}, \bibinfo{person}{Kevin Leyton-Brown}, {and} \bibinfo{person}{Yoav Shoham}.} \bibinfo{year}{2023}\natexlab{}.
\newblock \bibinfo{title}{In-context retrieval-augmented language models}.
\newblock
\newblock


\bibitem[Robertson et~al\mbox{.}(2009)]%
        {robertson2009probabilistic}
\bibfield{author}{\bibinfo{person}{Stephen Robertson}, \bibinfo{person}{Hugo Zaragoza}, {et~al\mbox{.}}} \bibinfo{year}{2009}\natexlab{}.
\newblock \showarticletitle{The probabilistic relevance framework: BM25 and beyond}.
\newblock \bibinfo{journal}{\emph{Foundations and Trends{\textregistered} in Information Retrieval}} \bibinfo{volume}{3}, \bibinfo{number}{4} (\bibinfo{year}{2009}), \bibinfo{pages}{333--389}.
\newblock


\bibitem[Roller et~al\mbox{.}(2021)]%
        {roller-etal-2021-recipes}
\bibfield{author}{\bibinfo{person}{Stephen Roller}, \bibinfo{person}{Emily Dinan}, \bibinfo{person}{Naman Goyal}, \bibinfo{person}{Da Ju}, \bibinfo{person}{Mary Williamson}, \bibinfo{person}{Yinhan Liu}, \bibinfo{person}{Jing Xu}, \bibinfo{person}{Myle Ott}, \bibinfo{person}{Eric~Michael Smith}, \bibinfo{person}{Y-Lan Boureau}, {and} \bibinfo{person}{Jason Weston}.} \bibinfo{year}{2021}\natexlab{}.
\newblock \showarticletitle{Recipes for Building an Open-Domain Chatbot}. In \bibinfo{booktitle}{\emph{Proceedings of the 16th Conference of the European Chapter of the Association for Computational Linguistics: Main Volume}}, \bibfield{editor}{\bibinfo{person}{Paola Merlo}, \bibinfo{person}{Jorg Tiedemann}, {and} \bibinfo{person}{Reut Tsarfaty}} (Eds.). \bibinfo{publisher}{Association for Computational Linguistics}, \bibinfo{address}{Online}, \bibinfo{pages}{300--325}.
\newblock
\urldef\tempurl%
\url{https://doi.org/10.18653/v1/2021.eacl-main.24}
\showDOI{\tempurl}


\bibitem[Salton and McGill(1983)]%
        {salton1983introduction}
\bibfield{author}{\bibinfo{person}{Gerard Salton} {and} \bibinfo{person}{Michael~J. McGill}.} \bibinfo{year}{1983}\natexlab{}.
\newblock \showarticletitle{Introduction to modern information retrieval}.
\newblock \bibinfo{journal}{\emph{McGraw-Hill}} (\bibinfo{year}{1983}).
\newblock


\bibitem[Santilli and Rodolà(2023)]%
        {santilli2023camoscio}
\bibfield{author}{\bibinfo{person}{Andrea Santilli} {and} \bibinfo{person}{Emanuele Rodolà}.} \bibinfo{year}{2023}\natexlab{}.
\newblock \bibinfo{title}{Camoscio: an Italian Instruction-tuned LLaMA}.
\newblock
\newblock
\showeprint[arxiv]{2307.16456}~[cs.CL]


\bibitem[Sauchuk et~al\mbox{.}(2022)]%
        {sauchuk2022role}
\bibfield{author}{\bibinfo{person}{Artsiom Sauchuk}, \bibinfo{person}{James Thorne}, \bibinfo{person}{Alon Halevy}, \bibinfo{person}{Nicola Tonellotto}, {and} \bibinfo{person}{Fabrizio Silvestri}.} \bibinfo{year}{2022}\natexlab{}.
\newblock \showarticletitle{On the Role of Relevance in Natural Language Processing Tasks}. In \bibinfo{booktitle}{\emph{Proceedings of the 45th International ACM SIGIR Conference on Research and Development in Information Retrieval}}. \bibinfo{publisher}{ACM}, \bibinfo{address}{Madrid}, \bibinfo{pages}{1785--1789}.
\newblock


\bibitem[Shazeer(2019)]%
        {shazeer2019mqa}
\bibfield{author}{\bibinfo{person}{Noam Shazeer}.} \bibinfo{year}{2019}\natexlab{}.
\newblock \bibinfo{title}{Fast Transformer Decoding: One Write-Head is All You Need}.
\newblock
\newblock
\showeprint[arxiv]{1911.02150}~[cs.NE]


\bibitem[Sun et~al\mbox{.}(2021)]%
        {sun2021long}
\bibfield{author}{\bibinfo{person}{Simeng Sun}, \bibinfo{person}{Kalpesh Krishna}, \bibinfo{person}{Andrew Mattarella-Micke}, {and} \bibinfo{person}{Mohit Iyyer}.} \bibinfo{year}{2021}\natexlab{}.
\newblock \bibinfo{title}{Do long-range language models actually use long-range context?}
\newblock
\newblock


\bibitem[Team et~al\mbox{.}(2023)]%
        {mosaicml2023introducing}
\bibfield{author}{\bibinfo{person}{MosaicML~NLP Team} {et~al\mbox{.}}} \bibinfo{year}{2023}\natexlab{}.
\newblock \bibinfo{title}{Introducing mpt-7b: A new standard for open-source, ly usable llms}.
\newblock
\newblock


\bibitem[Tolomei et~al\mbox{.}(2023)]%
        {prompt2os}
\bibfield{author}{\bibinfo{person}{Gabriele Tolomei}, \bibinfo{person}{Cesare Campagnano}, \bibinfo{person}{Fabrizio Silvestri}, {and} \bibinfo{person}{Giovanni Trappolini}.} \bibinfo{year}{2023}\natexlab{}.
\newblock \showarticletitle{Prompt-to-OS {(P2OS):} Revolutionizing Operating Systems and Human-Computer Interaction with Integrated {AI} Generative Models}. In \bibinfo{booktitle}{\emph{5th {IEEE} International Conference on Cognitive Machine Intelligence, CogMI 2023, Atlanta, GA, USA, November 1-4, 2023}}. \bibinfo{publisher}{{IEEE}}, \bibinfo{pages}{128--134}.
\newblock
\urldef\tempurl%
\url{https://doi.org/10.1109/COGMI58952.2023.00027}
\showDOI{\tempurl}


\bibitem[Touvron et~al\mbox{.}(2023a)]%
        {touvron2023llama}
\bibfield{author}{\bibinfo{person}{Hugo Touvron}, \bibinfo{person}{Thibaut Lavril}, \bibinfo{person}{Gautier Izacard}, \bibinfo{person}{Xavier Martinet}, \bibinfo{person}{Marie-Anne Lachaux}, \bibinfo{person}{Timoth{\'e}e Lacroix}, \bibinfo{person}{Baptiste Rozi{\`e}re}, \bibinfo{person}{Naman Goyal}, \bibinfo{person}{Eric Hambro}, \bibinfo{person}{Faisal Azhar}, {et~al\mbox{.}}} \bibinfo{year}{2023}\natexlab{a}.
\newblock \bibinfo{title}{Llama: Open and efficient foundation language models}.
\newblock
\newblock


\bibitem[Touvron et~al\mbox{.}(2023b)]%
        {touvron2023llama2}
\bibfield{author}{\bibinfo{person}{Hugo Touvron}, \bibinfo{person}{Louis Martin}, \bibinfo{person}{Kevin Stone}, \bibinfo{person}{Peter Albert}, \bibinfo{person}{Amjad Almahairi}, \bibinfo{person}{Yasmine Babaei}, \bibinfo{person}{Nikolay Bashlykov}, \bibinfo{person}{Soumya Batra}, \bibinfo{person}{Prajjwal Bhargava}, \bibinfo{person}{Shruti Bhosale}, {et~al\mbox{.}}} \bibinfo{year}{2023}\natexlab{b}.
\newblock \bibinfo{title}{Llama 2: Open foundation and fine-tuned chat models}.
\newblock
\newblock


\bibitem[Trappolini et~al\mbox{.}(2023)]%
        {mmndb}
\bibfield{author}{\bibinfo{person}{Giovanni Trappolini}, \bibinfo{person}{Andrea Santilli}, \bibinfo{person}{Emanuele Rodol{\`{a}}}, \bibinfo{person}{Alon~Y. Halevy}, {and} \bibinfo{person}{Fabrizio Silvestri}.} \bibinfo{year}{2023}\natexlab{}.
\newblock \showarticletitle{Multimodal Neural Databases}. In \bibinfo{booktitle}{\emph{Proceedings of the 46th International {ACM} {SIGIR} Conference on Research and Development in Information Retrieval, {SIGIR} 2023, Taipei, Taiwan, July 23-27, 2023}}, \bibfield{editor}{\bibinfo{person}{Hsin{-}Hsi Chen}, \bibinfo{person}{Wei{-}Jou~(Edward) Duh}, \bibinfo{person}{Hen{-}Hsen Huang}, \bibinfo{person}{Makoto~P. Kato}, \bibinfo{person}{Josiane Mothe}, {and} \bibinfo{person}{Barbara Poblete}} (Eds.). \bibinfo{publisher}{{ACM}}, \bibinfo{pages}{2619--2628}.
\newblock
\urldef\tempurl%
\url{https://doi.org/10.1145/3539618.3591930}
\showDOI{\tempurl}


\bibitem[Vaswani et~al\mbox{.}(2017)]%
        {vaswani2017attention}
\bibfield{author}{\bibinfo{person}{Ashish Vaswani}, \bibinfo{person}{Noam Shazeer}, \bibinfo{person}{Niki Parmar}, \bibinfo{person}{Jakob Uszkoreit}, \bibinfo{person}{Llion Jones}, \bibinfo{person}{Aidan~N Gomez}, \bibinfo{person}{\L~ukasz Kaiser}, {and} \bibinfo{person}{Illia Polosukhin}.} \bibinfo{year}{2017}\natexlab{}.
\newblock \showarticletitle{Attention is All you Need}. In \bibinfo{booktitle}{\emph{Advances in Neural Information Processing Systems}}, \bibfield{editor}{\bibinfo{person}{I.~Guyon}, \bibinfo{person}{U.~Von Luxburg}, \bibinfo{person}{S.~Bengio}, \bibinfo{person}{H.~Wallach}, \bibinfo{person}{R.~Fergus}, \bibinfo{person}{S.~Vishwanathan}, {and} \bibinfo{person}{R.~Garnett}} (Eds.), Vol.~\bibinfo{volume}{30}. \bibinfo{publisher}{Curran Associates, Inc.}, \bibinfo{address}{Long Beach}.
\newblock
\urldef\tempurl%
\url{https://proceedings.neurips.cc/paper_files/paper/2017/file/3f5ee243547dee91fbd053c1c4a845aa-Paper.pdf}
\showURL{%
\tempurl}


\bibitem[V{\"o}lske et~al\mbox{.}(2017)]%
        {reddit}
\bibfield{author}{\bibinfo{person}{Michael V{\"o}lske}, \bibinfo{person}{Martin Potthast}, \bibinfo{person}{Shahbaz Syed}, {and} \bibinfo{person}{Benno Stein}.} \bibinfo{year}{2017}\natexlab{}.
\newblock \showarticletitle{{TL};{DR}: Mining {R}eddit to Learn Automatic Summarization}. In \bibinfo{booktitle}{\emph{Proceedings of the Workshop on New Frontiers in Summarization}}, \bibfield{editor}{\bibinfo{person}{Lu~Wang}, \bibinfo{person}{Jackie Chi~Kit Cheung}, \bibinfo{person}{Giuseppe Carenini}, {and} \bibinfo{person}{Fei Liu}} (Eds.). \bibinfo{publisher}{Association for Computational Linguistics}, \bibinfo{address}{Copenhagen, Denmark}, \bibinfo{pages}{59--63}.
\newblock
\urldef\tempurl%
\url{https://doi.org/10.18653/v1/W17-4508}
\showDOI{\tempurl}


\bibitem[Wang et~al\mbox{.}(2024)]%
        {wang2024oop}
\bibfield{author}{\bibinfo{person}{Shuai Wang}, \bibinfo{person}{Liang Ding}, \bibinfo{person}{Li Shen}, \bibinfo{person}{Yong Luo}, \bibinfo{person}{Bo Du}, {and} \bibinfo{person}{Dacheng Tao}.} \bibinfo{year}{2024}\natexlab{}.
\newblock \showarticletitle{OOP: Object-Oriented Programming Evaluation Benchmark for Large Language Models}.
\newblock \bibinfo{journal}{\emph{arXiv preprint arXiv:2401.06628}} (\bibinfo{year}{2024}).
\newblock


\bibitem[Xie et~al\mbox{.}(2023)]%
        {xie2023adaptive}
\bibfield{author}{\bibinfo{person}{Jian Xie}, \bibinfo{person}{Kai Zhang}, \bibinfo{person}{Jiangjie Chen}, \bibinfo{person}{Renze Lou}, {and} \bibinfo{person}{Yu Su}.} \bibinfo{year}{2023}\natexlab{}.
\newblock \showarticletitle{Adaptive chameleon or stubborn sloth: Revealing the behavior of large language models in knowledge conflicts}. In \bibinfo{booktitle}{\emph{The Twelfth International Conference on Learning Representations}}.
\newblock


\bibitem[Xie et~al\mbox{.}(2024)]%
        {xie2024osworld}
\bibfield{author}{\bibinfo{person}{Tianbao Xie}, \bibinfo{person}{Danyang Zhang}, \bibinfo{person}{Jixuan Chen}, \bibinfo{person}{Xiaochuan Li}, \bibinfo{person}{Siheng Zhao}, \bibinfo{person}{Ruisheng Cao}, \bibinfo{person}{Toh~Jing Hua}, \bibinfo{person}{Zhoujun Cheng}, \bibinfo{person}{Dongchan Shin}, \bibinfo{person}{Fangyu Lei}, \bibinfo{person}{Yitao Liu}, \bibinfo{person}{Yiheng Xu}, \bibinfo{person}{Shuyan Zhou}, \bibinfo{person}{Silvio Savarese}, \bibinfo{person}{Caiming Xiong}, \bibinfo{person}{Victor Zhong}, {and} \bibinfo{person}{Tao Yu}.} \bibinfo{year}{2024}\natexlab{}.
\newblock \bibinfo{title}{OSWorld: Benchmarking Multimodal Agents for Open-Ended Tasks in Real Computer Environments}.
\newblock
\newblock
\showeprint[arxiv]{2404.07972}~[cs.AI]


\bibitem[Yates et~al\mbox{.}(2021)]%
        {yates-etal-2021-pretrained}
\bibfield{author}{\bibinfo{person}{Andrew Yates}, \bibinfo{person}{Rodrigo Nogueira}, {and} \bibinfo{person}{Jimmy Lin}.} \bibinfo{year}{2021}\natexlab{}.
\newblock \showarticletitle{Pretrained Transformers for Text Ranking: {BERT} and Beyond}. In \bibinfo{booktitle}{\emph{Proceedings of the 2021 Conference of the North American Chapter of the Association for Computational Linguistics: Human Language Technologies: Tutorials}}, \bibfield{editor}{\bibinfo{person}{Greg Kondrak}, \bibinfo{person}{Kalina Bontcheva}, {and} \bibinfo{person}{Dan Gillick}} (Eds.). \bibinfo{publisher}{Association for Computational Linguistics}, \bibinfo{address}{Online}, \bibinfo{pages}{1--4}.
\newblock
\urldef\tempurl%
\url{https://doi.org/10.18653/v1/2021.naacl-tutorials.1}
\showDOI{\tempurl}


\bibitem[Zhai et~al\mbox{.}(2023)]%
        {zhai2023stabilizing}
\bibfield{author}{\bibinfo{person}{Shuangfei Zhai}, \bibinfo{person}{Tatiana Likhomanenko}, \bibinfo{person}{Etai Littwin}, \bibinfo{person}{Dan Busbridge}, \bibinfo{person}{Jason Ramapuram}, \bibinfo{person}{Yizhe Zhang}, \bibinfo{person}{Jiatao Gu}, {and} \bibinfo{person}{Joshua~M Susskind}.} \bibinfo{year}{2023}\natexlab{}.
\newblock \showarticletitle{Stabilizing transformer training by preventing attention entropy collapse}. In \bibinfo{booktitle}{\emph{International Conference on Machine Learning}}. PMLR, \bibinfo{publisher}{PMLR}, \bibinfo{address}{Hawaii}, \bibinfo{pages}{40770--40803}.
\newblock


\bibitem[Zhan et~al\mbox{.}(2021)]%
        {zhan2021adore}
\bibfield{author}{\bibinfo{person}{Jingtao Zhan}, \bibinfo{person}{Jiaxin Mao}, \bibinfo{person}{Yiqun Liu}, \bibinfo{person}{Jiafeng Guo}, \bibinfo{person}{Min Zhang}, {and} \bibinfo{person}{Shaoping Ma}.} \bibinfo{year}{2021}\natexlab{}.
\newblock \showarticletitle{Optimizing Dense Retrieval Model Training with Hard Negatives}. In \bibinfo{booktitle}{\emph{Proceedings of the 44th International ACM SIGIR Conference on Research and Development in Information Retrieval}} \emph{(\bibinfo{series}{SIGIR '21})}. \bibinfo{address}{New York, NY, USA}, \bibinfo{pages}{1503–1512}.
\newblock
\showISBNx{9781450380379}


\bibitem[Zuccon et~al\mbox{.}(2023)]%
        {zuccon_hallucinates}
\bibfield{author}{\bibinfo{person}{Guido Zuccon}, \bibinfo{person}{Bevan Koopman}, {and} \bibinfo{person}{Razia Shaik}.} \bibinfo{year}{2023}\natexlab{}.
\newblock \showarticletitle{ChatGPT Hallucinates when Attributing Answers}. In \bibinfo{booktitle}{\emph{Proceedings of the Annual International ACM SIGIR Conference on Research and Development in Information Retrieval in the Asia Pacific Region}} (Beijing, China) \emph{(\bibinfo{series}{SIGIR-AP '23})}. \bibinfo{publisher}{Association for Computing Machinery}, \bibinfo{address}{New York, NY, USA}, \bibinfo{pages}{46–51}.
\newblock
\showISBNx{9798400704086}
\urldef\tempurl%
\url{https://doi.org/10.1145/3624918.3625329}
\showDOI{\tempurl}


\end{thebibliography}

\appendix

\end{document}